\documentclass{JHEP3}
\usepackage{graphicx,amsmath,amssymb}
\usepackage{epsfig,multicol}
\usepackage{epsf}
\usepackage{epstopdf}
\input{epsf.sty}
\newcommand{\D}{\overline{\mbox{D}}}
\newcommand{\e}{\epsilon}

\def\beq{\begin{equation}}
\def\eeq{\end{equation}}
\def\beqa{\begin{eqnarray}}
\def\eeqa{\end{eqnarray}}
\def\ba{\begin{eqnarray}}
\def\ea{\end{eqnarray}}
\def\be{\begin{equation}}
\def\ee{\end{equation}}

\def\D{\bar D}

\def\ap{\alpha^{\prime}}

\title{Comparing Brane Inflation to WMAP}
\author{Rachel Bean$^{1}$\footnote{rbean@astro.cornell.edu}~, 
Sarah~E. Shandera$^{2}$\footnote{sarah@phys.columbia.edu}~,
S.-H. Henry Tye$^3$\footnote{tye@lepp.cornell.edu}~ 
and Jiajun Xu$^3$\footnote{jx33@cornell.edu} 
\\ \small{\em $^1$ Department of Astronomy,
Cornell University, Ithaca, NY 14853}
\vskip .1cm
\\ \small{\em $^2$ Institute of Strings, Cosmology and Astroparticle Physics \\
Physics Department, Columbia University, New York, NY 10027}
\vskip .1cm
\\ \small{\em $^3$Newman Laboratory for Elementary Particle Physics \\
Cornell University, Ithaca, NY 14853}
}

\abstract{
We compare the simplest realistic brane inflationary model to recent cosmological data, including WMAP 3-year cosmic microwave background (CMB) results, Sloan Digital Sky Survey luminous red galaxies (SDSS LRG) power spectrum data and Supernovae Legacy Survey (SNLS) Type 1a supernovae distance measures.
Here, the inflaton is simply the position of a $D3$-brane which is moving towards a $\bar{D}3$-brane sitting at the bottom of a throat (a warped, deformed conifold) in the flux compactified bulk in Type IIB string theory. The analysis includes both the usual slow-roll scenario and the Dirac-Born-Infeld scenario of slow but relativistic rolling. Requiring that the throat is inside the bulk greatly restricts the allowed parameter space. We discuss possible scenarios in which large tensor mode and/or non-Gaussianity may emerge. Here, the properties of a large tensor mode deviate from that in the usual slow-roll scenario, providing a possible stringy signature. Overall, within the brane inflationary scenario, the cosmological data is providing information about the properties of the compactification of the extra dimensions.}

\begin{document}

\maketitle

\section{Introduction}

By now, inflation \cite{Guth:1980zm,Linde:1981mu,Albrecht:1982wi} 
is well established by the observational cosmic microwave background radiation (CMB) data
\cite{Smoot:1992td,Spergel:2006hy}. 
However, the inflationary universe is more like a paradigm than a theory, since a specific
compelling model is still missing. The best chance to come up with detailed working models is within a fundamental theory such as string theory.
String theory realizations of the inflationary universe may be separated into two types, depending on 
whether the inflaton is a closed or an open string mode. A prime example of the former type is 
when the closed string mode is a modulus, while for the latter type it is when 
the inflaton is the position of a brane inside the compactified bulk in a brane world. This is sometimes referred to as brane inflation \cite{Dvali:1998pa}, and the simplest specific realization as the KKLMMT scenario \cite{Kachru:2003sx}. We would like to make a detailed comparison of the predictions of this simple specific realization of brane inflation to the current cosmological data. This scenario offers potentially distinctive stringy signatures that may be detected. 

Brane inflation proposes a string theory motivated mechanism for inflation that appears to be generic in well-studied models which dynamically compactify the extra dimensions in a consistent way. It can have the additional nice feature of a natural ending when the branes collide, where the collision itself is useful for reheating and the possible production of cosmic strings. String theory dictates both the dynamics of the inflaton and its potential, so that one can make precise cosmological predictions from a given set of background parameters. Furthermore, given the potential of current and future measurements of the power spectrum including tensor mode perturbation, non-Gaussianities and possibly cosmic strings, the string theory parameters can eventually be over-constrained by the data \cite{Tye:2006uv}. The calculability and limited number of parameters make brane inflation an interesting arena to explore the possibilities for cosmology in string theory. 

In this paper, we make full use of current data to constrain the model. The particular brane inflationary model studied here is a simple realistic scenario with the inflaton mass as one of the free parameters. As our understanding of the compactification improves, one can probe numerous variations as well as more detailed features of the model. For this analysis we compare the model to the likelihood space allowed by Wilkinson Microwave Anisotropy Probe (WMAP) 3 year CMB data \cite{Spergel:2006hy,Hinshaw:2006ia,Jarosik:2006ib,Page:2006hz} in combination with the Sloan Digital Sky Survey luminous red galaxies (SDSS LRG) matter power spectrum \cite{Tegmark:2006az} and the Supernovae Legacy Survey (SNLS) supernovae sample \cite{Astier:2005qq}. 

Consistency checks will provide strong limits on input parameters, and we can further narrow the choice using the measurement of the spectral index $n_s$, the tensor/scalar ratio $r$, and current limits on non-Gaussianity. However, this data still leaves us with a few distinct regions of parameter space that we would like to examine in more detail. We outline below the basic differences between usual slow-roll and the brane inflation scenario, our results for calculating simple cosmological observables, and the features we expect in different regions of the input parameter space. If brane inflation is correct, one may hope that data will rule out everywhere except for a point (or a tiny region) in the whole parameter space. Hopefully, this work also gives a clear idea how further theoretical analysis, together with more data, will likely over-constrain the model and provide a critical test of the brane inflationary scenario.
Within the brane inflationary scenario, the cosmological data relates directly to properties of the compactification of the extra dimensions. In contrast to today, where the standard model branes are sitting at a corner of the bulk, the mobile branes in the early universe probe a different part of, and complementary information about, the bulk space. That the inflaton is an open string mode and so takes the Dirac-Born-Infeld kinetic form in its effective action leads to interesting new features \cite{Silverstein:2003hf}.

We shall present our analysis in three steps:
\begin{enumerate}
 \item We re-examine the analysis in the literature and find agreement of the model with data for some  ranges of the inflaton mass $m$, which is treated as a free parameter here. There is (i) the small 
 $m$ region \cite{Kachru:2003sx,Firouzjahi:2005dh,Seljak:2006hi}, which is essentially the KKLMMT model, (ii) the intermediate mass region, where the tensor mode can be very large and which deviates from the usual slow-roll relation between the size of the tensor mode perturbation and its spectral index $n_{t}$
 \cite{Shandera:2006ax},
and (iii) the large mass region, where non-Gaussianity (due to the DBI action) may be large and detectable \cite{Silverstein:2003hf,Alishahiha:2004eh}.
We find there are regions with a red tilted power spectrum, in agreement with the data in each of these cases.
\item We consider the impact of requiring self-consistency, that the throat should be inside the compactified manifold, which must be finite for the Newton's constant to be non-vanishing. Requiring the size of the throat to be smaller than the volume of the bulk imposes a very strong condition on the size of the throat and so on how far the $D$3-brane can be away from the bottom of the throat. This condition, which is absent in field theory models, rules out eternal inflation of the random walk type in brane inflation \cite{Chen:2006hs}. This bulk volume bound also limits the e-folds allowed.  In generic situations, imposing this bound here rules out both large tensor and large non-Gaussianity in the model \cite{Baumann:2006th}. This leads to the conclusion that the original KKLMMT scenario (with massless inflaton, very small tensor and non-Gaussianity) seems to fit the data best.

\item We then discuss how the bulk volume bound may be satisfied in a number of variations of the simplest brane inflation model. As specific possibilities, we discuss a couple of scenarios where the large tensor mode may be present. Since this tensor mode would have a spectrum that deviates from that in the slow-roll scenario, this may be a way to pick up a string theory signature. These scenarios also allow large non-Gaussianity.
\end{enumerate}

The paper proceeds as follows: in section \ref{model} we outline the brane inflation model. 
In section \ref{cosmo} the background and perturbation evolution and the dependency of power spectrum observables on brane/throat parameters are discussed. Four different inflationary regimes in the DBI model are discussed.
Section \ref{WMAPcomp} presents the constraints on the inflationary model in light of cosmological observations. 
In section \ref{largetensor}, we discuss a specific scenario where the tensor mode can be large.
In section \ref{conclusion} we tie together our findings and discuss implications for the future. Unless otherwise specified, mass units are given in term of the reduced Planck mass.

\section{The brane inflation model}\label{model}

Suppose our universe today is described by a brane world solution in Type IIB string theory. In this scenario, 6 of the 9 spatial dimensions are dynamically compactified to a finite size. 
There are $D$3-branes (and probably $D$5- and $D$7-branes as well) that span our observable universe. Standard model particles are the lightest open string modes which must end on branes.
The $D$3-branes are point-like in the 6-dimensional compactified manifold known as the bulk. Closed strings can be everywhere, with the graviton being the lightest mode.
This flux compactification that is dynamically stabilized automatically introduces a warped geometry, with regions in this bulk known as warped (deformed) throats \cite{Giddings:2001yu,Kachru:2003aw}. 

In the early universe, it is easy to imagine additional $D$3-${\D}$3-brane pairs around. A $\D$3-brane has the same tension but opposite Ramond-Ramond (R-R) charge as a $D$3-brane, and is attracted (strongly) to the bottom of a throat. Inflation takes place while the mobile $D$3-brane is moving towards the ${\D}$3-brane and inflation ends when they collide and annihilate each other 
\cite{Burgess:2001fx,Dvali:2001fw,Shandera:2003gx}. Fluctuations that are present before inflation, such as defects, radiation or matter, will be inflated away. Here, the $D$3-brane position $\phi$ is the inflaton and the inflaton potential $V(\phi)$ comes from the brane tensions and interactions. The $D$3-${\D}$3-brane annihilation releases the brane tension energy that heats up the universe to start the hot Big Bang epoch. 
So brane inflation relies on the dynamics of the $D$3-brane extended in our usual space-time and moving in the extra dimensions.  
Together, the 10-dimensional metric takes the form:
\be
ds^2=h^{2}(r)(-dt^{2}+a(t)^{2} d{\bf x}^{2})
+ h^{-2}(r)(dr^2+r^2 d\Sigma_{X_5}^2),
\label{10dmetric}
\ee
Here the cosmic scale factor $a(t)$ is that of an expanding homogeneous isotropic universe spanned by the 3-dimensions ${\bf x}$, 
and $r$ is the coordinate along the throat. The warp factor $h(r)$ and the metric on the extra dimensions ($d\Sigma^2_{X_5}$) are inputs to the model. Warped spaces are natural in string theory models and are useful for flattening potentials and for generating a hierarchy of scales with the UV at the top (edge) of the throat and the IR scale at the warped bottom (around $r \sim r_{A}$). We will use one particularly well-motivated form for the full metric with $h(r)\sim r/R$, where $R \gg r_{A}$ is the scale of the throat. The throat will be described by these two numbers giving its length and warping. The inflaton $\phi$ is related to the position of a space-time filling $D$3-brane moving in such a throat. Specifically, 
\be
\phi=\sqrt{T_3}r
\ee 
where $T_3=[(2\pi)^3g_s\alpha^{\prime2}]^{-1}$ is the tension of a $D$3-brane. Here $g_{s}$ is the string coupling and the Regge slope $\alpha^{\prime}=m_{s}^{-2}$ sets the string scale, where $m_{s}$ is the string mass scale.

String theory suggests not only the metric for this model, but also the appropriate action and potential. 
Besides a topological piece (the Chern-Simons term), it is well-known that the world volume action for a $D$-brane also involves the Dirac-Born-Infeld (DBI) action, which has an unusual kinetic term \cite{Fradkin:1985qd,Abouelsaood:1986gd}.  Here we will use a particularly simple background geometry, so that the action is correspondingly simple 
\cite{Silverstein:2003hf} :
\be
S=-\int d^4x\;a^3(t)\left[T\sqrt{1- \dot{\phi}^2/T} + V(\phi) - T \right]
\label{DBIact}
\ee
where 
\be
T(\phi) = T_{3}h^4(\phi)
\label{warpedtension}
\ee 
is the warped $D$3-brane tension at $\phi$ and $V(\phi)$ is the inflaton potential. 
The DBI action has been used in string cosmology before \cite{Kehagias:1999vr,Burgess:2003qv}.
The important new ingredient here is the combination of warped geometry with the DBI action, a consequence of the dynamical moduli stabilization in flux compactification. 
To see the key point, it is natural to introduce the Lorentz factor, $\gamma$, that tracks the motion of the brane :
\be
\label{gamma1}
\gamma(\phi)=\frac{1}{\sqrt{1-\dot{\phi}^2/T(\phi)}}
\ee
Note that the inflaton speed is limited by the warped tension, i.e., $\dot{\phi}^2 < T(\phi)$ and $T(\phi)$ decreases towards the bottom of the throat, irrespective of the steepness of the inflaton potential. As the limiting speed is decreasing rapidly down the throat, this can lead to a scenario where the inflaton is moving slowly but ultra-relativistically, generating predictions that are quite different from those of the usual slow-roll case. In the non-relativistic limit, $\gamma \rightarrow 1$, the model (\ref{DBIact}) reduces to those with a standard canonical kinetic term.

While the form above is useful for the calculations that follow, it is worthwhile to display the more complete and fundamental expression for the action\cite{Polchinski:1998rr}. For interesting and straightforward generalizations of the background geometry there can be many fields appearing non-trivially in the action, including the dilaton $\Phi$, the metric $G_{\mu\nu}$, the anti-symmetric tensor $B_{\mu\nu}$ and the gauge field $F_{\mu\nu}$. The Chern-Simons term contains couplings between the brane and R-R fields ($p$-forms) $C_p$, with $p$ even for type IIB theory. We use variable $\xi$ and indices $\{a,b\}$ for coordinates and quantities on the brane. Then, ignoring the potential which we will discuss in detail below, we have
\ba
\label{fullaction}
S_{D3}&=&-T_3\int d^4\xi e^{-\Phi}\:\sqrt{det |G_{ab}+B_{ab}+2\pi\ap F_{ab}|}\\\nonumber
&&\pm \mu_3\int_{\mathcal{M}_4}\left[\sum_{p=0}^4C_{p}\right]\wedge Tr \left[e^{2\pi\ap F+B}\right]
\ea
where the quantities inside the determinant have been pulled back onto the brane and the $\pm$ is for a brane/anti-brane. In particular, note that the pulled back metric ($G_{ab}$) will contain the warp factor. In the second term $\mu_3=g_sT_3$ is the brane charge. The integration is over the $D$3-brane world volume, and contributions should, of course, have the correct dimension. In the simplest cases, many of the fields appearing in (\ref{fullaction}) are trivially zero. However, for more interesting solutions we will need to solve the supergravity equations to find each of the fields. 

\begin{figure}[h]
\begin{center}
\leavevmode
\includegraphics[width=0.7\textwidth,angle=0]{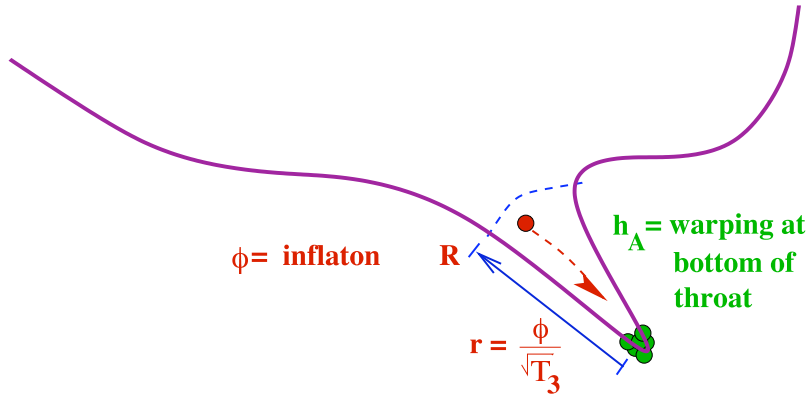} 
\caption{A cartoon of the throat in the compact extra dimensions. There may in general be warped throats of a variety of sizes attached to the bulk space. $R$ sets the scale of the throat, while $h_A$ is the warping at the bottom. The $D$3-brane moves down the throat, attracted by a $\bar{D}$3-brane (or a stack of them) sitting at the bottom. The inflation, $\phi$, is related to the brane position, $r$.
\label{throatfig}}
\end{center}
\end{figure}

We see now that the simple case in (\ref{DBIact}) has constant dilaton (set to 0), vanishing pullbacks $B_{ab}$ and $F_{ab}$, and only $C_4$ non-zero ($D$3-branes are charged under the 4-form RR field $C_{4}$). The $D$3-brane alone in this background is supersymmetric. In addition, the supergravity equations require that components of $C_4$ on the brane be $h^4(r)/g_s$. We have aligned the brane coordinates with the usual space-time coordinates so that the only non-zero derivatives in the pullback are those with respect to time, and we assume that only the radial motion of the brane is important. Then (\ref{fullaction}) simplifies to
\be
S_{D3}=\int d^4x\:a^3(t)T_3 h^4(r) \left(-\sqrt{1-h^{-4}\dot{r}^2} + 1 \right)
\ee
From the supergravity solution, we find that the throat should be characterized by the parameters 
$h_A=h(r_{A})$, the warp factor at the bottom, of the throat, where $r=r_{A}$, and a background number of charges we label $N_A$. The scale of the throat $R$ is given by \cite{Gubser:1998vd}
\be
\label{throatR}
R^4=4\pi g_sN_A\alpha^{\prime2}/{v}
\ee 
where $v$ is the ratio of the dimensionless volume of the space $X_5$ (in (\ref{10dmetric})) to that of a 5-sphere (which has volume $\pi^3$, assuming unit radius). For the well studied Klebanov-Strassler (KS) throat, which has a simple geometry (to be discussed in section \ref{KST}), $v=16/27$ \cite{Gubser:1998vd}. However, the KS throat has an $S^{2}$ symmetry, so one can consider its orbifolded versions, where $v$ can be quite small ($\ll1$). 
We sketch the throat in Fig.\ref{throatfig}.

The inflaton potential can in principle be computed from a detailed knowledge of the string theory background (the flux compactification) and is of the form
\be
V(\phi)=\frac{1}{2}m^2\phi^2+V_0+V_{D\bar{D}}+\dots
\label{inflatonpot}
\ee
Let us first consider the constant and brane/anti-brane interaction terms, which in more detail are
(for $\gamma \simeq 1$),
\be
V_0+V_{D\bar{D}}=2T_3h_A^4\left(1-\frac{1}{N_{A}}\frac{r_A^4}{r^4}\right)=V_0\left(1-\frac{V_0}{4\pi^2v}\frac{1}{\phi^4}\right)
\ee
The constant term, providing the vacuum energy for inflation, comes from the warped tension of the brane and anti-brane. The interaction term comes from the attraction between the mobile $D$3-brane and a $\D$3-brane sitting naturally at the bottom (small $r$ end) of the throat. Its form can be understood in several ways, most simply as the analog of the Coulomb interaction in the 6 dimensions transverse to the branes. The interaction can be explicitly derived in flat space by examining closed string exchange between the branes, and the calculation can be extended for relativistic branes to determine $\gamma$ factor corrections (see (\ref{potential1})). Finally, in the warped geometries considered here, one brane can be treated as a probe moving in the compact space with a metric perturbed by the other brane \cite{Kachru:2003sx}. This explains the factor of $1/N_A$, since the perturbation due to a single extra brane should be suppressed by the number of background branes. Note that the singularity of the Coulomb term is never reached. When the branes are within the string scale distance, the Coulomb potential is curtailed due to the exchange of the infinite tower of closed string modes, as well as the emergence of a tachyon mode, whose rolling signifies the collision and annihilation of the $D$3-$\D$3 brane pair.

So we have
\begin{equation} \phi_{e} \ge \phi_{i } > \phi_{pivot} > \phi_{E} > \phi_{A}\nonumber\end{equation}
where $\phi_{e}$ is the value at the edge of the throat, $\phi_{i}$ is at the onset of inflation, $\phi_{pivot}$ is the position when the observed CMB scale was originally produced, $\phi_{E}$ is when the tachyon emerges and inflation ends, and $\phi_{A}$  at the bottom of the throat.

While the constant and the coefficient of the Coulombic term are easily calculated in most cases, finding the effective mass, $m$, is quite involved \cite{McAllister:2005mq,Shandera:2004zy}. Contributions from the Kahler potential and superpotential after moduli stabilization may be large \cite{Kachru:2003sx}, and some work on simple models indicates that $\ap$ corrections are also important \cite{Berg:2004ek,Berg:2004sj,Balasubramanian:2004uy}. In realistic examples, calculating $m$ is a long and cumbersome task (although the situation is improved somewhat by the techniques of \cite{Baumann:2006th}). Because of these difficulties, we take the pragmatic approach of considering $m$ to be a parameter of the model. The cosmological predictions are very sensitive to the value of $m^{2}$ and actually the shape of the throat \cite{Baumann:2006th,Burgess:2006cb}. In fact, the size of $m$ partially dictates how closely this model mimics standard slow-roll and where it deviates. Furthermore, as shown in 
Ref.\cite{Firouzjahi:2005dh}, even a small $m^{2}$ in the slow-roll case modifies the power spectrum from red tilt to blue tilt.
Since the value of $m^{2}$ is dictated by the details of the compactified manifold, CMB data could reveal to us the structures of the manifold, a very exciting aspect. There may also be other terms in the potential. For large values of $\phi$, we should include additional terms like $\phi^{4}$, which we will discuss later in section \ref{WMAPcomp}.  

Cosmic strings are generically produced during the brane annihilation towards the end of inflation.
Although we shall not discuss the cosmological implications of cosmic strings here, we shall occasionally quote the prediction of the cosmic string tension $\mu$ in terms of the dimensionless parameter $G \mu$. 

\section{Cosmological implications of brane inflation}\label{cosmo}

In this section we outline the background expansion and the spectrum of density and tensor perturbations for the DBI inflation model.

\subsection{Homogeneous background evolution}

There is an important difference between the DBI scenario and the more usual slow-roll models often presented, which shows up in the kinetic term for the inflaton. The important feature of the DBI action (\ref{DBIact}) is the effective speed limit imposed by the Lorentz factor $\gamma$ (\ref{gamma1}) 
introduced above \cite{Kabat:1999yq, Silverstein:2003hf}. 

Notice that the speed limit depends on the warping, so that deep in the throat (small $r$ and $\phi$), $\gamma$ may grow quickly. This effect allows the model to achieve enough e-folds of inflation even in a steep potential and is the source of several interesting observable differences from the standard slow-roll. We emphasize that this behavior is an unavoidable feature, coming from the fact that the brane position is an open string mode.

Using the action in (\ref{DBIact}) modifies the usual cosmology \cite{Silverstein:2003hf}. The pressure $p$ and density $\rho$ are given by
\ba
\label{prho}
\rho=&T(\gamma-1)+V \\\nonumber
p=&T(\gamma - 1)/\gamma - V
\ea
The equation of motion for the inflaton is
\be
\label{exacteom}
\ddot{\phi}-\frac{3T^{\prime}}{2T}\dot{\phi}^2+T^{\prime}+\frac{3H}{\gamma^2}\dot{\phi}+\left(V^{\prime}-T^{\prime}\right)\frac{1}{\gamma^3}=0
\ee
where $\gamma$, $H$, and $V$ are all functions of $\phi$ and the prime denotes derivatives with respect to $\phi$. We have used the continuity equation
\be
\dot{\rho}=-3H(\rho+p).
\ee
Notice that for small velocity, (\ref{prho}) and (\ref{exacteom}) reduce to the usual expressions.  
Some algebra gives a useful expression for $\dot{\phi}$
\be
\label{phidot}
\dot{\phi}=\frac{-2H^{\prime}}{\sqrt{1/M_p^4+4T^{-1}H^{\prime2}}}
\ee
We emphasize that we have not neglected the ${\ddot{\phi}}$ term. From the Friedmann equations, we need $w=p/\rho<-1/3$. 
Using (\ref{prho}),
\be
\frac{\ddot{a}}{a} = \frac{V}{3M_{p}^{2}} - \frac{T(\gamma +2 -3/\gamma)}{6M_{p}^{2}}
\ee
where the kinetic term always contributes negatively. 
Then
\be
\label{cosm}
3M_p^2H(\phi)^2=V(\phi)+T(\gamma(\phi)-1)
\ee
Using the solution to the equation of motion for $\phi$, we may write
\be
\label{gamma2}
\gamma(\phi) = \sqrt{1+4M_p^4T^{-1}H^{\prime}(\phi)^2}
\ee
Physically, this model has a sound speed that changes with $\gamma$ as
\be
c_s=\frac{1}{\gamma}
\ee
The sound speed plays a role in the scalar power spectrum, discussed below, and in the shape of non-Gaussianities \cite{Chen:2006nt}. Equations (\ref{cosm}) and (\ref{gamma2}) give a differential equation for $H(\phi)$. Once the Hubble parameter and $\gamma(\phi)$ are known, one can find expressions for the power spectrum observables. We use the following parameters:
\ba
\label{params1}
\epsilon&\equiv&\frac{2M_p^2}{\gamma}\left(\frac{H^{\prime}(\phi)}{H(\phi)}\right)^2=-\frac{\dot{H}}{H^2}\\\nonumber
\eta&\equiv&\frac{2M_p^2}{\gamma}\left(\frac{H^{\prime\prime}(\phi)}{H(\phi)}\right)\\\nonumber
\kappa&\equiv&\frac{2M_p^2}{\gamma}\left(\frac{H^{\prime}(\phi)}{H(\phi)}\frac{\gamma^{\prime}(\phi)}{\gamma}\right)=\frac{\dot{c_s}}{c_sH}\\\nonumber
\tilde{\eta}&\equiv&\frac{\dot{\epsilon}}{\epsilon H}=2\epsilon-2\eta+\kappa
\ea
The far right hand sides and $\tilde{\eta}$ are for comparison with \cite{Baumann:2006cd} and \cite{Chen:2006nt}. Note that inflation takes place for $0<\epsilon <1$, since
\be
\label{0eps1}
\frac{\ddot{a}}{a}=H^2(1-\epsilon)
\ee
All three parameters must stay less than one for the expressions for the observables to be valid. In the non-relativistic limit $\gamma \rightarrow 1$, they are simply related to the usual slow-roll parameters $\epsilon_{SR}$ and $\eta_{SR}$: $\epsilon \rightarrow \epsilon_{SR}$, $\eta \rightarrow \eta_{SR}- \epsilon_{SR}$ and $\kappa \rightarrow 0$, where $\epsilon_{SR} \equiv \frac{1}{2}(V'/V)^2$ and $\eta_{SR} \equiv V''/V$.

\subsection{Scalar and tensor perturbations}

The scalar density perturbation has been studied in 
Ref.\cite{Garriga:1999vw,Alishahiha:2004eh,Stewart:1993bc}.
Decomposing the inflaton into its rolling background $\phi(t)$ and a fluctuation $\delta$,
\ba
\phi= \phi(t) + \delta(x,t)
\ea
with a scalar perturbation (the Newtonian potential) $\Phi$ in the de-Sitter metric,
\be
ds^{2} = -(1 + 2 \Phi)dt^{2} +a(t)^{2}(1 - 2 \Phi)dx^{2}
\ee
it is easy to see that $\delta$ and $\Phi$ are related so that there is only one independent scalar fluctuation. The linear combination
$$\zeta = \frac{H}{\dot \phi} \delta + \Phi$$
becomes frozen as it exits the horizon during inflation, later generating the temperature fluctuation 
in the cosmic microwave background radiation. The evolution of $\zeta$ obeys a linearized Einstein equation. In terms of the variable 
\ba
\label{zdef}
z = \frac{a{\dot \phi} \gamma^{3/2}}{H}
\ea
one introduces the scalar density perturbation $u=\zeta z$. Introducing the conformal time $\tau$, 
$d\tau = dt/a$, we see that $u_{k}$ as a function of the wavenumber $k$ satisfies  
\ba
\label{ukeq}
\frac{d^{2}u_{k}}{d\tau^{2}} + \left( \frac{k^{2}}{\gamma^{2}} - \frac{1}{z} \frac{d^{2}z}{d\tau^{2}} \right) u_{k}=0
\ea
Note that the fluctuations $u$ travels at the sound speed $c_{s}$,
\ba
c_{s}^{2}= \frac{\partial p}{\partial \dot \phi} / \frac{\partial \rho}{\partial \dot \phi} = \frac{1}{\gamma^{2}}
\ea
so $u_{k}$ freezes when $k$ crosses $k=aH\gamma$ instead of $k=aH$. To solve for $u_{k}$, we express the potential in terms of inflationary properties. 
Following (\ref{zdef}), we obtain
\ba
\label{d2zdt}
\frac{1}{z} \frac{d^{2}z}{d\tau^{2}} &=&  a^{2}H^{2} W  \\\nonumber
W &=&  2(1+\epsilon-\eta-\frac{\kappa}{2})
(1- \frac{\eta}{2} - \frac{\kappa}{4}) 
- \epsilon_{N} +\eta_{N} + \frac{\kappa_{N}}{2}
\ea 
where we introduce the derivative of the DBI parameters with respect to the e-fold number $N$, e.g.,
$\eta_{N}=\eta,_{N}= \frac{d\eta}{dN}$,
\ba
\frac{\dot \phi}{H} \epsilon^{\prime} &=& - \epsilon_N = \epsilon( 2  \epsilon -2 \eta + \kappa) \\ \nonumber 
\frac{\dot \phi}{H} \eta^{\prime} &=& - \eta_{N}= \eta(\epsilon +  \kappa) - \xi  \\ \nonumber
\frac{\dot \phi}{H} \kappa^{\prime} &=& - \kappa_{N}= \kappa (2 \kappa +\epsilon - \eta) - \epsilon\omega
\ea
where
\ba
\xi&\equiv&\frac{4M_p^4}{\gamma^2}\left(\frac{H^{\prime}(\phi)H^{\prime\prime\prime}(\phi)}{H^2(\phi)}\right)\\\nonumber
\omega&\equiv&\frac{2M_p^2}{\gamma}\left(\frac{\gamma^{\prime\prime}}{\gamma}\right)
\ea
Expressing the Hubble scale in conformal time (valid when $\epsilon$ is roughly constant), $aH\tau(1-\epsilon)= -1$,
(\ref{ukeq}) becomes
\ba
\frac{d^{2}u_{k}}{d\tau^{2}} + \left( c_{s}^{2}k^{2} - \frac{\nu^{2}-1/4}{\tau^{2}} \right) u_{k}=0
\ea
where
$$\nu^{2} = \frac{W}{(1-\epsilon)^{2}} + \frac{1}{4}$$
Here, $\nu \rightarrow 3/2$ as the DBI parameters vanish. Since all DBI parameters, as well as $H$, vary much more slowly than $a(t)$, we may take $\nu$ to be close to constant, so the above equation behaves as a Bessel equation. We see that, for $aH \gamma \gg k$, the growing mode behaves as
\ba
|u_{k}| \rightarrow 2^{\nu - 2} \frac{\Gamma(\nu)}{\Gamma(\frac{3}{2})} \frac{1}{\sqrt{c_sk}} (c_sk\tau)^{1/2 -\nu} 
\ea
so the spectral density is given by
\ba
\label{spectralden} 
{\mathcal P}_R^{1/2}(k) = \sqrt{\frac{k^{3}}{2 \pi^{2}}} \left|\frac{u_{k}}{z}\right|
= 2^{\nu - 3/2} \frac{\Gamma(\nu)}{\Gamma(\frac{3}{2})} (1-\epsilon)^{\nu -1/2} \left.\frac{H^2}{2\pi |\dot{\phi}|}\right|_{k=aH\gamma}
\ea
Then the scalar spectral index $n_{s}$ is given by 
\ba
n_s -1& \equiv & \frac{d\ln{\mathcal P}_R}{d\ln\:k} \nonumber \\
&\simeq& (1+\epsilon+\kappa)( - 4\epsilon + 2\eta - 2\kappa)\label{nsm1} 
\ea
which reduces to the usual slow-roll formula $n_s -1=  - 6\epsilon_{SR} + 2\eta_{SR} $ in the limit $\gamma\rightarrow 1$. As the $D$3-brane moves down the throat, $\gamma, \kappa$ and $\epsilon$ tend to increase. In the large $\gamma $ limit in the AdS throat, $\kappa \rightarrow - 2\epsilon $ 
while $\eta  \rightarrow 0$. Then $n_s-1\rightarrow 0$.

The tensor mode spectral density, to first order, is given by
\ba\label{tensor}
{\mathcal P}_h=\left.\frac{2H^2}{M_p^2\pi^2} \right|_{k=aH}
\ea
and the corresponding tensor power index:
\beq
n_t \equiv \frac{d\ln\:{\mathcal P}_h}{d\ln\:k} \approx \frac{-2\epsilon}{1-\epsilon-\kappa}
\eeq
This is non-vanishing even in the ultra-relativistic case. The ratio of power in tensor modes versus scalar modes is
\be
\label{tsratio}
r=\frac{16\epsilon}{\gamma}
\ee
To keep $r \lesssim0.5$, we would like $\gamma$ to increase as $\epsilon$ does. However, the
 non-Gaussianity bound constrains $\gamma \lesssim 31$. This bound can be saturated under certain conditions in the intermediate regime. In the intermediate regime, it is possible to have 
 $\epsilon\sim 0.2$ and $\gamma\sim1$, so that $r$ exceeds the current bound.

Note that these equations suggest a key way in which one may distinguish usual slow-roll inflation from the DBI scenario: the consistency relationship between the tensor power index and the tensor/scalar ratio is modified
\be\label{ntgamma}
n_t=-\frac{r}{8}\left(\frac{\gamma}{1-\epsilon-\kappa}\right)
\ee
This is in fact a common feature of non-standard inflation models with sound speed $c_s$ less than 1 \cite{Garriga:1999vw}. To first order, $n_t=-r/(8c_s)$. Since $\gamma$ is greater than one (and grows with time), finding the magnitude of $n_t$ larger than $r/8$ would be evidence for a non-slow-roll scenario. It was pointed out in Ref.\cite{Alishahiha:2004eh} that the non-Gaussianity $f_{NL}$ is proportional to $\gamma^2$. The present observational bound from non-Gaussianity $|f_{NL}|  \simeq 0.32 \gamma^{2} \lesssim 300$ yields $\gamma \lesssim 31$, where the limit on non-Gaussianities coming from DBI-type models is discussed in Ref.\cite{Chen:2006nt, Creminelli:2005hu}. The calculation of the full bispectrum for models of inflation with small sound speed (including DBI) is done in Ref.\cite{Chen:2006nt}. The trispectrum is studied in Ref.\cite{Huang:2006eh}. We note that the four observables $f_{NL}$, $r$, $n_t$, and $n_s$ can be related to the four parameters $\gamma$, $\epsilon$, $\kappa$ and $\eta$.

\subsection{Impact of throat warping: The Klebanov-Strassler Throat}\label{KST}

The power spectrum index is sensitive to the deformation of the throat once we are away from the slow-roll region, and as such the precise shape of the throat, i.e., the warped geometry and the deformation can be measured using observations \cite{Shiu:2006kj}.
For this reason we discuss the warped throat in some detail here. 

Consider two different approximate ways to incorporate the deformation of the warped conifold, namely the {\it AdS cut-off} and the {\it mass gap} (MG) cases for which the D3-brane tension is given by:
\ba
T_{AdS}(\phi) &=& \frac{\phi^{4}}{\lambda},  \quad   \phi \ge \phi_{A} \nonumber \\
T_{MG}(\phi) &=& \frac{(\phi^{2}+b^{2})^{2}}{\lambda},  \quad \phi \ge 0 \\
\lambda &\equiv& T_3 R^4 = \frac{27}{16} \frac{N_{A}}{2\pi^2}  \nonumber
\ea
We expect $\phi_{A} \simeq b$, so away from the bottom of the throat, for large $\phi$, we expect little difference in physics. However, in Ref.\cite{Shandera:2006ax}, we find $n_{s} -1$ is slightly positive ($\sim \epsilon^{2}/\gamma$) while in the second case, Ref.\cite{Kecskemeti:2006cg} finds that the spectrum has a red tilt. This difference appears in the DBI region. We expect a small (hopefully observable) effect of the warp factor in the large mass region. (The warp factor drops out in the slow-roll region.)
This means $n_{s}$ and its running will tell us about the shape of the throat, an exciting prospect. 
Note that the location and size of the 4-cycles in the bulk (partially) measures the inflaton mass \cite{Baumann:2006th}. So we really can learn a lot about the compactification geometry from the CMB.

Now we have to consider the geometry of the throat more carefully. 
Although there is a wide class of possible geometries, we only know the full metric of the Klebanov-Strassler (KS) throat. So let us consider this case. Consider the 10d-metric (\ref{10dmetric}),
\ba
\label{metric}
ds^{2} &=& h(r)^{2}(g_{{\mu \nu}}d^{\mu}xd^{\nu}x) +  h(r)^{-2}ds_{6}^{2}  \nonumber \\
&=&h(r)^{2}(-dt^{2}+a^{2}(t)d{\bf x}^{2}) +  h(r)^{-2}(dr^{2} + r^2 d \Sigma^2) \nonumber \\
&=& h(\tau)^{2} (-dt^{2}+a^{2}(t)d{\bf x}^{2}) +  h(\tau)^{-2}(\frac{\epsilon^{4/3}}{6K^{2}(\tau)} 
d\tau^{2} + ...)
\ea
(here, $\epsilon$ is the deformation parameter, not the inflationary parameter) where
\ba
\label{K(tau)}
K(\tau)=\frac{(\, \sinh(2\tau)-2\tau\,)^{1/3}}{2^{1/3}\sinh \tau} \, 
\ea
Unless we want to study the iso-curvature density perturbations, we shall ignore the shape 
encoded in $d \Sigma^2$. (For slow-roll, we expect this to be small, but they may be not in the DBI case.) 
The warp factor $h(\tau)$ is given by the following integral expression \cite{Klebanov:2000hb}
\ba
\label{h(tau)}
h^{-4}(\tau)= 2^{2/3}\,(g_s M \alpha')^2\, \epsilon^{-8/3}\, I(\tau)\, ,
\ea
\ba
\label{I(tau)}
I(\tau)\equiv\int_{\tau}^{\infty} d\, x \frac{x\coth\, x -1}{\sinh^2 \,x }
\left(\,\sinh(2x)-2x\, \right)^{1/3} \, .
\ea
where $M$ is a parameter characterizing the background flux. Both $K(\tau)$ and $I(\tau)$ are well-behaved functions of $\tau$:
\ba
\label{asym}
K(\tau \rightarrow 0)\rightarrow \, (2/3)^{1/3}(1 +\frac{\tau^{2}}{30}+ . . .), \quad
K(\tau \rightarrow \infty)\rightarrow \, 2^{1/3}\, e^{-\tau/3}\nonumber\\
I(\tau \rightarrow 0)\rightarrow \, a_0 +{\cal{O}}(\tau^2), \quad
I(\tau \rightarrow \infty)\rightarrow \, 3\, .\, 2^{-1/3} 
\left(\tau-\frac{1}{4} \right) e^{-4\tau/3}
\ea
where $a_0 \sim 0.71805$. Here we take $\epsilon^{2/3}$ to have dimension of length.
So we have 
\ba
h(r) &=& h(\tau)  \nonumber\\
dr &=& \frac{\epsilon^{2/3}}{\sqrt{6}K(\tau)} d\tau
\ea
Integrating this equation gives the relation between the throat position,$r$, and $\tau$,
\be
\label{rtaurelate}
r-r_{0} = \frac{1}{2^{5/6}3^{1/6}}\epsilon^{2/3} \tau \left(1 + \frac{\tau^{2}}{18} + ... \right)
\ee
From the explicit expressions in terms of $\tau$, we can obtain $h(r)$ or $T(\phi)$.
For large $\tau$, its relation to $r$ in  $ds_6^2$ is given by
\ba
\label{r}
r^{2} \simeq \frac{3}{2^{5/3}}\e^{4/3}e^{2\tau/3}
\ea
In the limit $\e \to 0$, finite $r$ implies $\tau \to \infty$, so
\ba
ds_{6}^{2} \to dr^{2} + r^{2} d \Sigma^2
\ea
which is simply the conifold metric.
At $\tau=0$, we have $h_{A}=h(r_{0})=r_{0}/R$ and
\be 
h(r_{0})^{2}=h(\tau=0)^{2}= c_{0}\frac{1}{g_{s}M} \frac{\epsilon^{4/3}}{\alpha'}= e^{-4 \pi K/3 g_{s}M}
\ee
where $c_{0}= 1/(2^{1/3}\sqrt{a_{0}})$, so
\be
r^{2}_{0} = c_{0}\frac{1}{g_{s}M}\left(\frac{\epsilon^{4/3}}{\alpha'}\right) R^{2}
\ee
where $R$ is given by (\ref{throatR}) and $N_{A}=KM$. 

The edge of the throat is glued smoothly to the bulk of the generalized Calabi-Yau (CY) manifold. The bulk of the CY manifold is defined to have no significant warping, so that the warp factor in the bulk is essentially of order unity. Suppose $\tau_c$ is the point where the throat is glued to the bulk. We may estimate $\tau_c$ using (\ref{rtaurelate}):
\begin{eqnarray}
\frac{\epsilon^{2/3}}{\sqrt{6}}\int_0^{\tau_c}\frac{1}{K(\tau)}d\tau&\approx& R-r_0\approx\frac{\epsilon^{2/3}}{\sqrt{6}}\int_0^{\tau_c}\frac{1}{2^{1/3}e^{-\tau/3}}d\tau\\\nonumber
\Rightarrow e^{\tau_c/3}&\approx&(R-r_0)\epsilon^{-2/3}\frac{2^{1/3}\sqrt{6}}{3}+1
\end{eqnarray}
as long as $\tau_c>>1$, which is true for reasonable parameter values. This expression agrees well with the numerical value for which $h(\tau_c)= 1$.

\begin{figure}[b]
\begin{center}
\includegraphics[width=0.6\textwidth,angle=0]{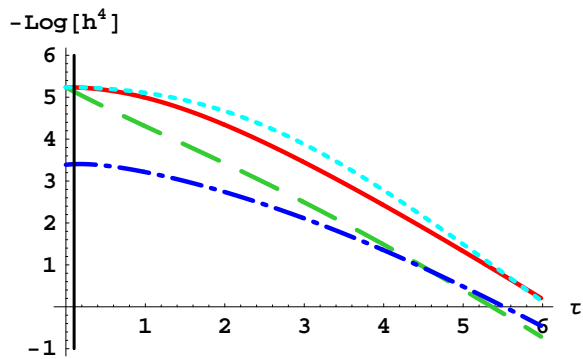} 
\caption{Comparison of the various warp factor expressions. The solid (red) line is the full warped deformed conifold expression. The long dashed (green) line is the AdS warp factor. The dot-dashed (dark blue) is the log-corrected expression, and the short-dashed (light blue) is the mass-gap. For this plot $N_A=10^6$, $M=4000$. The vertical black line indicates $\tau_E$, where the tachyon develops and inflation ends.}
\label{largeM}
\end{center}
\end{figure}

The relation between $\tau_c, \e$ and $l_s$ is important to impose the constraint that the throat must
fit inside the bulk. The physical size of the throat is given by
\ba
\label{throatl}
l=6^{-1/2}\, \e^{2/3}\, \int_0^{\tau_c} d\tau \, \frac{h^{1/4}(\tau)}{K(\tau)} \, 
\sim \sqrt{g_s M}\, l_s \, \tau_c
\ea
We have assumed that $l < V_6^{1/6}$, where $V_6$ is the volume of the compactification. 

\begin{figure}[b]
\begin{center}
\centering{
\hbox{\includegraphics[width=8.5cm]{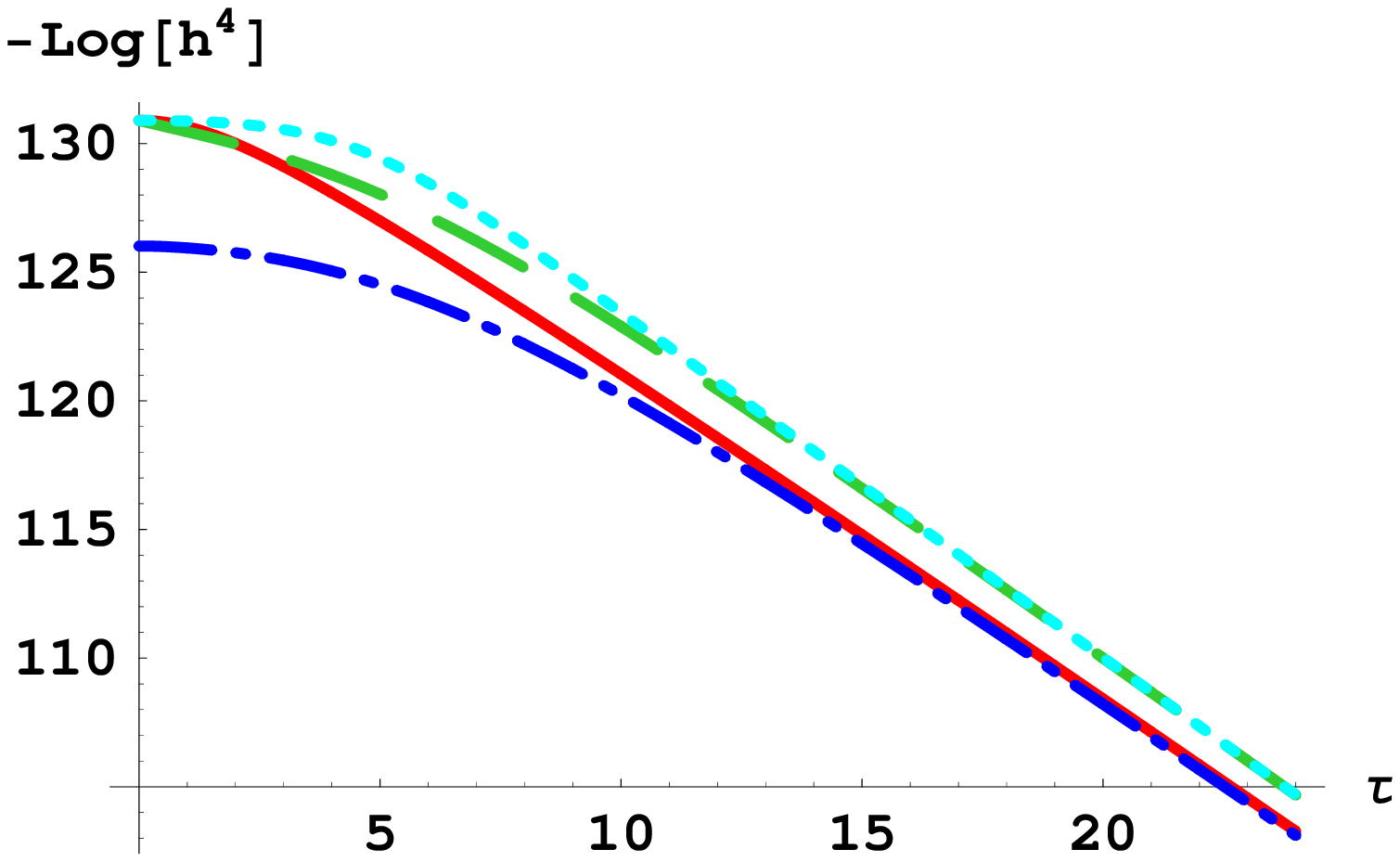}\includegraphics[width=8.5cm]{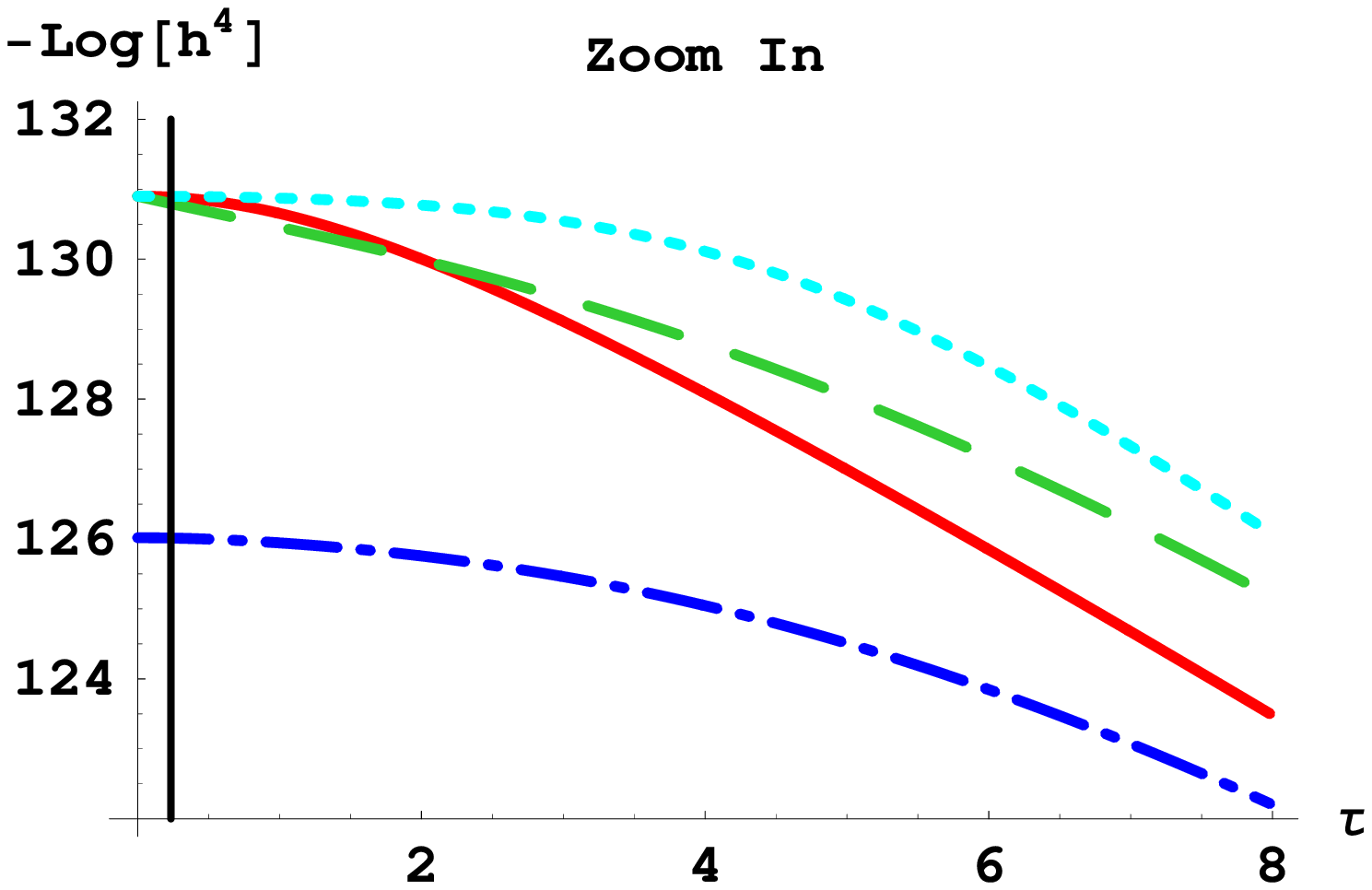}}
\hbox{\includegraphics[width=8.5cm]{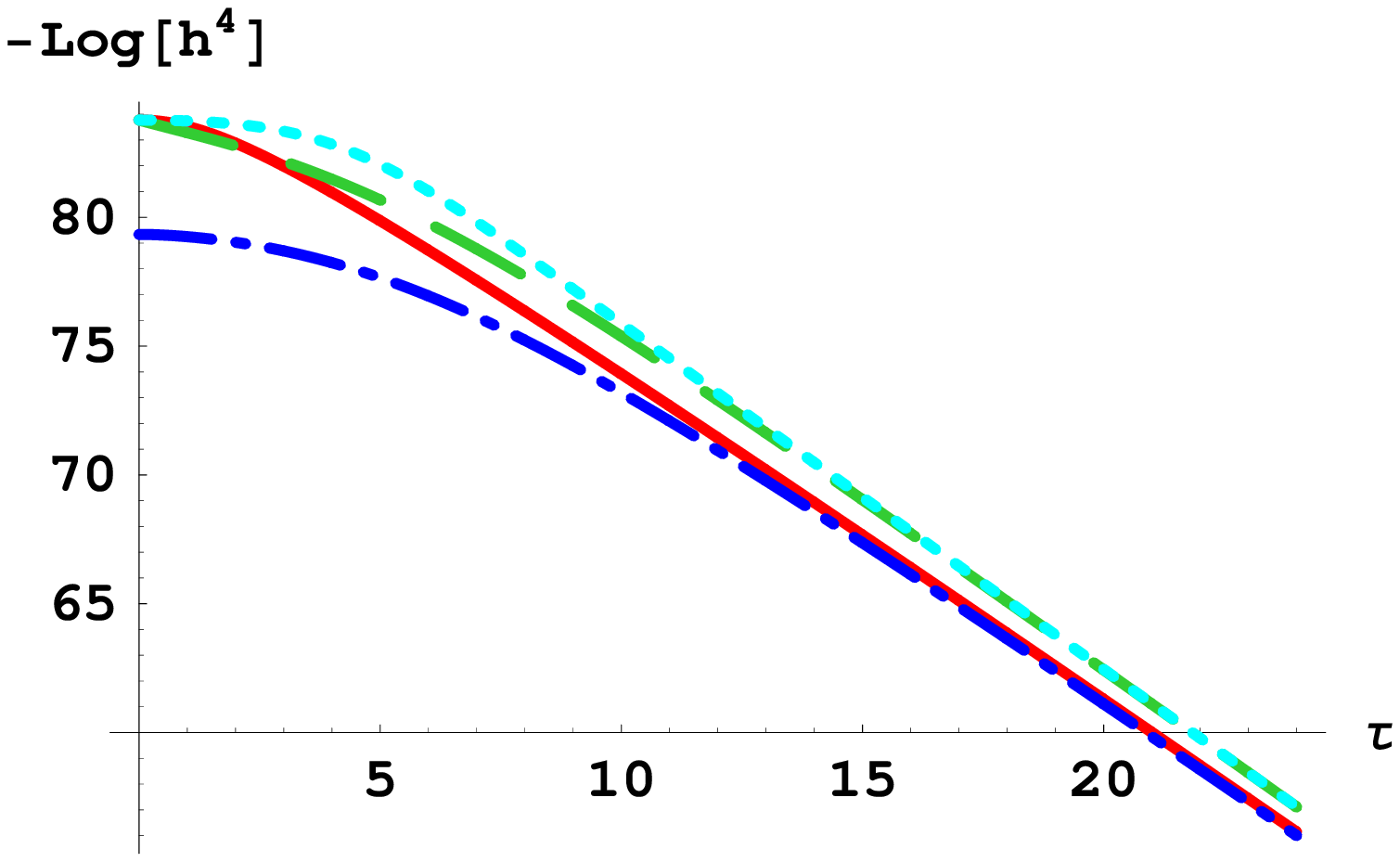}\includegraphics[width=8.5cm]{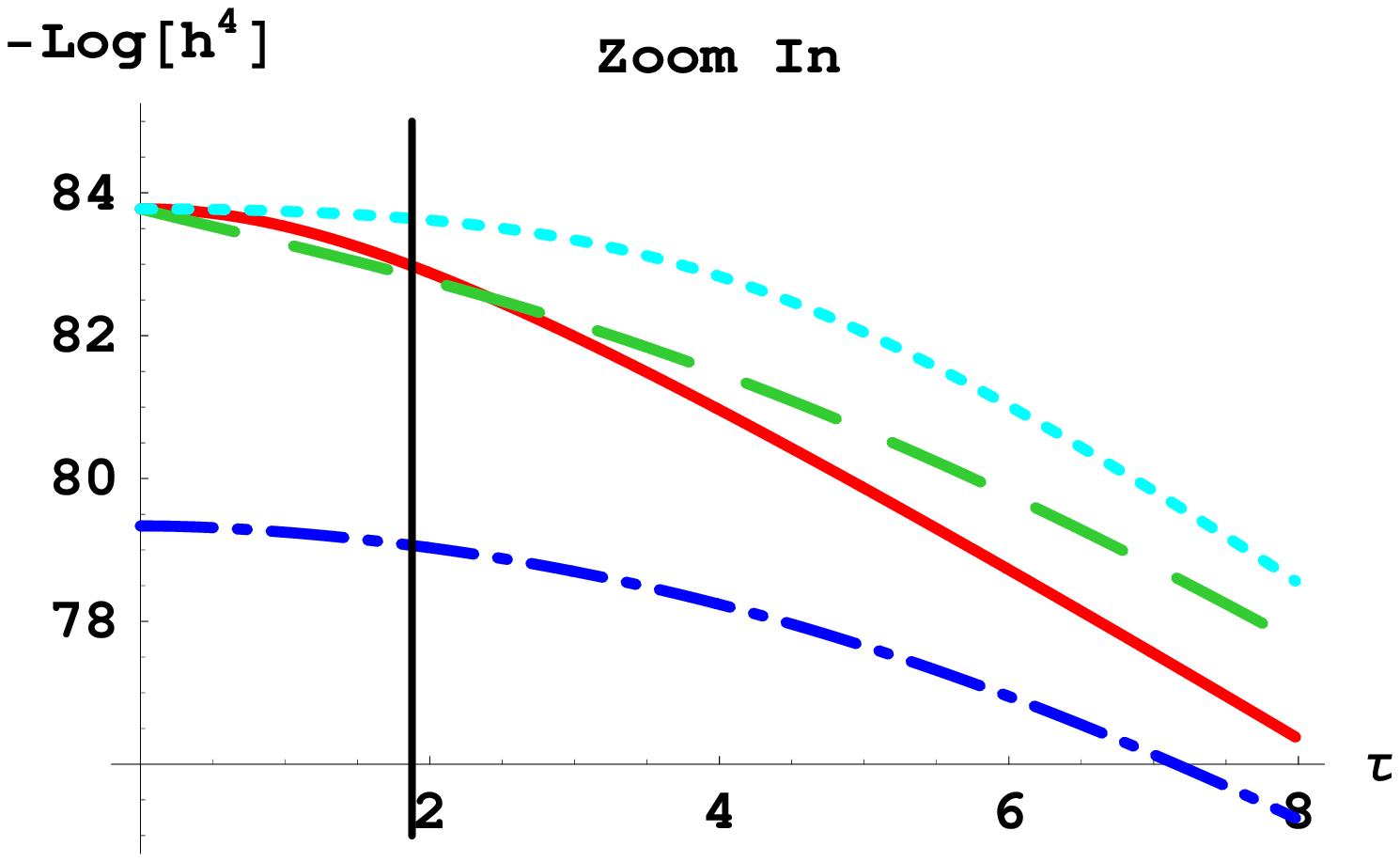}}
}
\caption{As in Figure \protect \ref{largeM} showing warp factors for (top panels) $N_A=10^6$, $M=800$ and (bottom panels) $N_A=10^2$, $M=10$. The right hand panels show a zoomed in region. The vertical black line indicates $\tau_E$, where the tachyon develops and inflation ends.}\label{smallfig}
\end{center}
\end{figure}

Which approximate warp factor is closer to the actual solution depends on details of how the throat is cut-off. To illustrate this point, we show numerical examples of the various expressions for the warp factor in Figures \ref{largeM} and \ref{smallfig}. Note that when $M>\sqrt{N_A}$, the mass-gap formula is much more accurate than the AdS expression. In particular, the AdS warp factor indicates a throat that is much too short (the point where the throat connects to the bulk is when $h(\tau_c)=1$). However, when $M<\sqrt{N_A}$, the mass-gap expression is actually too flat and the AdS expression is more accurate. This is especially true for small $N_A$, where the slope of the full solution changes considerably in one warped string length. For completeness, we also plot the log-corrected solution (see \cite{Klebanov:2000hb}), although we do not use it in our analysis since it deviates strongly from the actual KS solution at small $\tau$. 

One can understand these results by examining the initial slopes of the various expressions, accounting for the change of coordinates between $r$ and $\tau$ in the small $\tau$ limit. The important constant in the AdS and mass-gap expressions is $r_0/\epsilon^{2/3}\sim (\sqrt{N_A}/M)^{1/2}$. If this constant is large, the slope of the mass-gap expression decreases and that warp factor flattens out, while the AdS expression is less negative. If the constant is small, the AdS case has a very large negative slope initially, while the mass-gap case has a stronger $\tau$ dependence. 

\subsection{E-folds and initial power spectrum observables}
The observable power spectrum parameters are measured at a range of wave numbers, $k$, which in turn correspond to
a certain time range before the end of inflation. It is more convenient to translate this to the number of e-folds $N_e$ before the end of inflation.  
This number of e-folds is given by the integral 
\begin{equation}
N_e(\phi) = \int H\;dt = \int_{\phi_i}^{\phi}\frac{H(\phi)}{\dot{\phi}} d\phi = \frac{1}{2M_p^2} \int_{\phi}^{\phi_i}\frac{H(\phi)\gamma(\phi)}{H^\prime(\phi)} d\phi
\label{Nefolds}
\end{equation}
In order to calculate the e-fold number correctly, it is crucial to determine the position of the brane when inflation ends, i.e. $\phi_E$.  From (\ref{0eps1}), we know that inflation ends when $\epsilon > 1$.  But in most cases, $\epsilon$ remains less than one and inflation ends for a different reason.  In the slow-roll scenario, $\eta$ grows much faster than $\epsilon$ during inflation and it is actually $\eta = 1$ that sets $\phi_E$. For the DBI scenario, $\epsilon$ is suppressed by $1/\gamma$ and almost never grow larger than 1 at the end of inflation. In fact, in the DBI scenario, inflation ends when the proper distance between $D\bar{D}$ pairs is the string scale $\sqrt{\alpha'}$. 

Using the metric (\ref{10dmetric}), the proper distance between the branes can be calculated as
\begin{equation}\label{brane_distance}
\int ds = \int \frac{dr}{h(r)}.
\end{equation}
For the AdS throat, with $h(r) = r/R$, we get
\begin{equation}
\phi_E = \phi_A e^{\sqrt{\alpha'}/{R}}
\end{equation}
For the mass-gap warp factor, with $h(r) = \sqrt{r^2 + r_0^2}/R$, we find
\begin{equation}
\phi_E = \phi_0 \sinh \left( \frac{\sqrt{\alpha'}}{R}  \right)
\end{equation}
where $\phi_0=\sqrt{T_3}r_0$ and the anti-brane now sits at $\phi=0$. 

\subsection{Four inflationary scenarios}\label{threereg}
The full expression for the inflaton potential is dependent upon $\gamma$,
\be
\label{potential1}
V  = \frac{m^2}{2}\phi^2 +V_{0}\left(1-\frac{V_{0}}{4 \pi^{2} v \phi^4}\frac{(\gamma+1)^2}{4\gamma}\right)
\ee
 where the Coulomb term is important in the non-relativistic regime, $\gamma \simeq 1$, but does not play an important role when the inflaton is relativistic, $\gamma \gg 1$.

The inflaton mass is proportional to the Hubble scale so we can parameterize it as $m^2 \sim \beta H^2$. Now $\beta$ controls whether the mass term or the constant term $V_0$ is dominant in the potential and thus determines whether inflation is slow-roll or relativistic. In the slow-roll case, we also have $H^2 \sim V_0$, so that $\beta$ may be treated as a constant.  For the relativistic case, the potential is dominated by the mass term (so we no longer have $H^2 \sim V_0$) but we can still parameterize $m^2$ in terms of  the constant term $V_0$ as $m^2 \sim \beta V_0$. 

Overall, there are four scenarios, three for a $D$3-brane moving down a throat and one for a $D$3-brane coming out of a throat : 

\begin{enumerate}

\item  $\beta \ll 1$, $\gamma \simeq 1$, the {\it slow-roll} case, when $m^{2} \simeq 0$; 
the $m^{2}=0$ case is studied in Ref.\cite{Kachru:2003sx} and the more general small $m^{2}$ case is studied in Ref.\cite{Firouzjahi:2005dh,Seljak:2006hi}, where it was found that $\beta <0.05$; that is, the range $0.05 \lesssim \beta \lesssim 0.2$ is ruled out. Likewise they find, for small 
$\beta$, $n_{s} \sim 0.98 +\beta$, the tensor to scalar ratio $\log(r) \sim -8.8 +60\beta$, and the cosmic string tension $\log(G\mu) \sim -9.4 + 30 \beta$. As we shall see in our analysis, $n_s \simeq 0.97$ when $\beta \rightarrow 0$. To explain the difference here, one should note that when $\beta=0$, $n_s - 1 \sim -5/(3N_e)$ \cite{Kachru:2003sx}, and in our analysis, the minimum number of e-folds is not fixed, but rather depends on the scale of inflation through (\ref{min_efold}). So the allowed range of $\beta$ is now more subtle as it depends on $V_0$ which sets the inflation scale in the slow-roll case.

\item $\beta \sim 1$, $\gamma \simeq 1$ during inflation, but $\gamma$ increases to a large value towards the end of inflation; this corresponds to some {\it intermediate} values of $m^{2}$. This case offers the best hope of a large tensor mode that has a tensor power index $n_{t}$ different from that predicted by the slow-roll scenario \cite{Shandera:2006ax}. This scenario is not easily studied analytically and is more suited to numerical integration as we discuss further scenario in section \ref{WMAPcomp}.

We can see the origin of the large tensor modes by combining the differential form of the e-fold expression (\ref{Nefolds}) with (\ref{phidot}) to give the usual Lyth bound expression
\be
\label{Lyth}
\frac{1}{M_{P}}\frac{d\phi}{dN_{e}}=\frac{2M_{P}H^{\prime}}{\gamma H} =\sqrt{\frac{r}{8}}
\ee
where $r$ is a function of $\phi$. For small $r$, one can get many e-folds for a relatively small displacement of $\phi$. For relatively 
large $r$, it is difficult in usual slow-roll to get enough e-folds for an initial $\phi_{i}$ that is bounded (i.e., less than $M_P$). To agree with data, we see that large $r$ can be achieved during inflation only if $r$ drops rapidly as $\phi$ decreases, and inflation ends. This is achieved in the DBI model when $\gamma$ increases rapidly as the D3-brane moves down the throat ($r  \sim \gamma^{-1}$, since the parameter $\epsilon$ is nearly constant). This means that this intermediate scenario can have large $r$, but it is inevitably accompanied by large non-Gaussianity on smaller scales.

\item $\beta \gg 1$, $\gamma$ is large, inflation is {\it ultra-relativistic} throughout. This region typically has large distinctive non-Gaussianity \cite{Alishahiha:2004eh,Chen:2006nt} and has been further analyzed recently \cite{Shiu:2006kj}. We shall go into some details of this scenario in the next section.

\item For completeness, we would like to mention another interesting case, namely, when $\beta <0$, when the inflaton mass is tachyonic.
The scenario becomes the multi-throat brane inflation scenario proposed by Chen \cite{Chen:2004gc,Chen:2005ad}. The Coulombic term $V_{D \D}$ is negligible and inflation takes place as the $D$3-brane moves out of a throat (before it moves into a second throat). For small tachyonic mass, this is simply a slow-roll model. This DBI inflation can happen when inflaton mass takes a generic value, $|m| \approx H$.
The distance the inflaton travels through during inflation, $\Delta \phi
\approx H R^2 \sqrt{T_3}$, is always sub-Planckian.
This model may be realized in a multi-throat compactification starting
with a number of antibranes settled down at the ends of various throats.
These antibranes are classically stable, but can annihilate against
the fluxes quantum mechanically via tunneling \cite{Kachru:2002gs}, producing
the $D$3-branes, which then leave the throat. 
This model predicts large non-Gaussianity with the same shape as in the UV
model. The difference is the running, $f_{NL} \approx 0.036 \beta^2 N_e^2$, that is,
here $f_{NL}$ decreases with $k$, while,  in the UV model, $f_{NL}$ increases with $k$. 
The power spectrum index undergoes an interesting phase transition at a critical e-fold, 
from red ($n_s-1 \approx -4/N_e$) at small scales to blue ($n_s -1  \sim 4/N_e$) 
at large scales \cite{Chen:2005ad,Chen:2005fe}. 
If such a transition falls into the observable range of WMAP/PLANCK, it predicts a
large running of $n_s$, i.e., a large negative $dn_s/d\ln k$ in some region, but 
is un-observably small otherwise. 

\end{enumerate}

\subsection{Ultra-relativistic inflation}\label{ultrarel}

In the large $\gamma$ case, the inflaton mass $m$ is relatively big and one may neglect the other two terms in the potential (\ref{potential1}). 
To compare with the discussion in Ref.\cite{Silverstein:2003hf,Kecskemeti:2006cg}, let us choose the deformation to be given by a mass gap,
\be
f^{-1}(\phi)=T(\phi)= \frac{(\phi^{2}+b^{2})^{2}}{\lambda} \simeq \frac{b^{4}}{\lambda} + 
\frac{2b^{2}}{\lambda} \phi^{2}
\ee
where $b$ measures the deformation and $\lambda \sim N_{A}$.
The density perturbation 
\be 
P(k) \sim 10^{-10} = \frac{fV^{2}}{36 \pi^{2} M_{p}^{4}} = 
\frac{\lambda}{144 \pi^{2}} \left(\frac{\phi_{i}^{2}}{\phi_{i}^{2} + b^{2}}\right)^{2} 
\left(\frac{m}{M_{p}}\right)^{4}
\ee
and 
\ba 
N_{e} \simeq 55 &=& \int H \frac{d\phi}{\dot \phi} = \int \frac{d\phi}{M_{p}} \sqrt{fV/3} \nonumber \\
&=& \sqrt{\lambda/6}\frac{m}{2 M_{p}} \ln \left[ \frac{\phi_{i}^{2}+b^{2}}{\phi_{E}^{2}+b^{2}} \right]
\ea
so $\delta_{H} \simeq N_{e}^{2}/\sqrt{\lambda}$, yielding
\be
\label{mmpratio}
\lambda \simeq 10^{17} \qquad \frac{m}{M_{p}}\simeq 10^{-6}
\ee

In terms of the throat length $R$, the value of the inflaton at the edge of the throat is 
\ba
\phi_{e}=\sqrt{T_{3}}R \simeq m_{s} \lambda^{1/4} 
\ea
and 
\be
b = m_{s}^{2}Rh_{A}
\ee
For the SUGRA approximation to be valid, we require
\be
\label{mmsratio}
\frac{m^{2}}{2}\phi_{e}^{2} \lesssim m_{s}^{4} \quad \rightarrow \quad \frac{m}{m_{s}} \lesssim \lambda^{-1/4} \sim 10^{-4} 
\ee
Combining (\ref{mmpratio}) and (\ref{mmsratio}), we find
\be
\label{msmpr}
\frac{m_{s}}{M_{p}} \gtrsim 10^{-2}
\ee
Now we have, for non-Gaussianity,
\be
f_{NL} \simeq \left(\frac{m}{M_{p}}\right)^{2}\left(\frac{M_{p}}{m_{s}h_{A}}\right)^{4}
\simeq 10^{-12 }\frac{1}{(G \mu_{s})^{2}}
\ee
where $\mu_{s}$ is the cosmic string tension. For $G \mu_{s} < 10^{-7}$, $f_{NL}$ will exceed observational limits.
So the presence of large non-Gaussianity implies cosmic strings should be absent. This can happen if there is no $\D$3-brane at the bottom of the throat.

If a $\D$3-brane is at the bottom of the throat, the tachyon appears at $\phi_{t} = h_{A}m_{s}$, which is typically smaller than $b$. Also, we would like $h_{A} > 10^{-4}$ to ensure efficient (re-)heating \cite{Chen:2006ni}. This choice is consistent with the other conditions so far.

So far, we have not yet demanded that the throat must be inside the bulk. For typical throats that we know of, the width of the throat is comparable to its length, i.e., $R_{w} \simeq R$. If that is the case, then we find that 
\be
\label{sizeconst}
\frac{m_{s}}{M_{p}} \lesssim \lambda^{-3/4} \sim 10^{-12}
\ee
which is clearly incompatible with the condition (\ref{msmpr}). Seeing non-Gaussianity (and assuming a $D$3 brane inflaton), therefore, implies that the throat is not a KS throat or any of the $Y^{p,q}$ type. 
The $S^{2}$ in the KS throat has an $S^{2}$ symmetry, so one may choose to orbifold it. Generically, we expect the size of the edge of the throat is $R^{4}/p$ for ${\bf Z}_{p}$. For a large enough $p$, we may relax the above constraint (\ref{sizeconst}). This constraint translates to the bulk volume bound (\ref{Liambound}) introduced in Ref.\cite{Hailu:2006uj} since the inflaton $\phi$ is measured in units of $m_s$ and the bulk size (which determines $M_p$) can be compared to the known throat scale $R^4\propto N_A\alpha^{\prime2}$.

As we shall see, consistent with this discussion, when combined with data, imposing the bulk volume bound would rule out the region of the parameter space with large $\gamma$ (and large non-Gaussianity).

\section{Comparison with cosmological observations}\label{WMAPcomp}
In the previous section we outlined how  the initial spectrum of scalar and tensor perturbations arising from brane inflation can be characterized in terms of six parameters: the inflaton mass, $m$,  background charge, $N_A$, warping, $h_A=h(\phi_A)$, the orbifold/ volume ratio parameter in (\ref{throatR}), $v$,  string length, $\sqrt{\ap}$, and the string coupling, $g_s$. To compare with observations we consider the predicted spectrum normalization, $A_{s}(k_{pivot})\equiv P_{R}(k_{pivot})$, scalar and tensor spectral indexes, $n_{s}(k_{pivot})$ and $n_{t}(k_{pivot})$, and tensor to scalar ratio, $r = P_{h}(k_{pivot})/P_{R}(k_{pivot})$ , at a near-horizon physical scale $k_{pivot}=0.002/Mpc$. 

For the following numerical analyses, unless otherwise stated, we fix the string coupling $g_s= 0.1$ and length $\ap = 1000$. We investigate the constraints on the remaining four parameters \{$N_A, v, h_{A},m^{2}$\}, and the slow-roll, intermediate and ultra-relativistic regimes outlined in section \ref{threereg}. We generate Monte Carlo selected brane inflation models, obtained using a flat prior on \{$\log N_{A}, \log v, \log m^{2}, A_{s}(k_{pivot})$\} and finding $h_{A}$ to give the required value of $A_{s}(k_{pivot})$. As mentioned in section \ref{threereg}, although the slow-roll and ultra-relativistic regimes can be easily studied analytically,  the intermediate regime cannot. In appendix \ref{app1} we outline the numerical approach to integrating the equation of motion and perturbation equations, and in appendix \ref{app2} we describe how the inflationary initial power spectrum is translated to the physical scales today. 

\begin{figure}[t]
\begin{center}
\leavevmode
\includegraphics[height=0.7\textheight,angle=0]{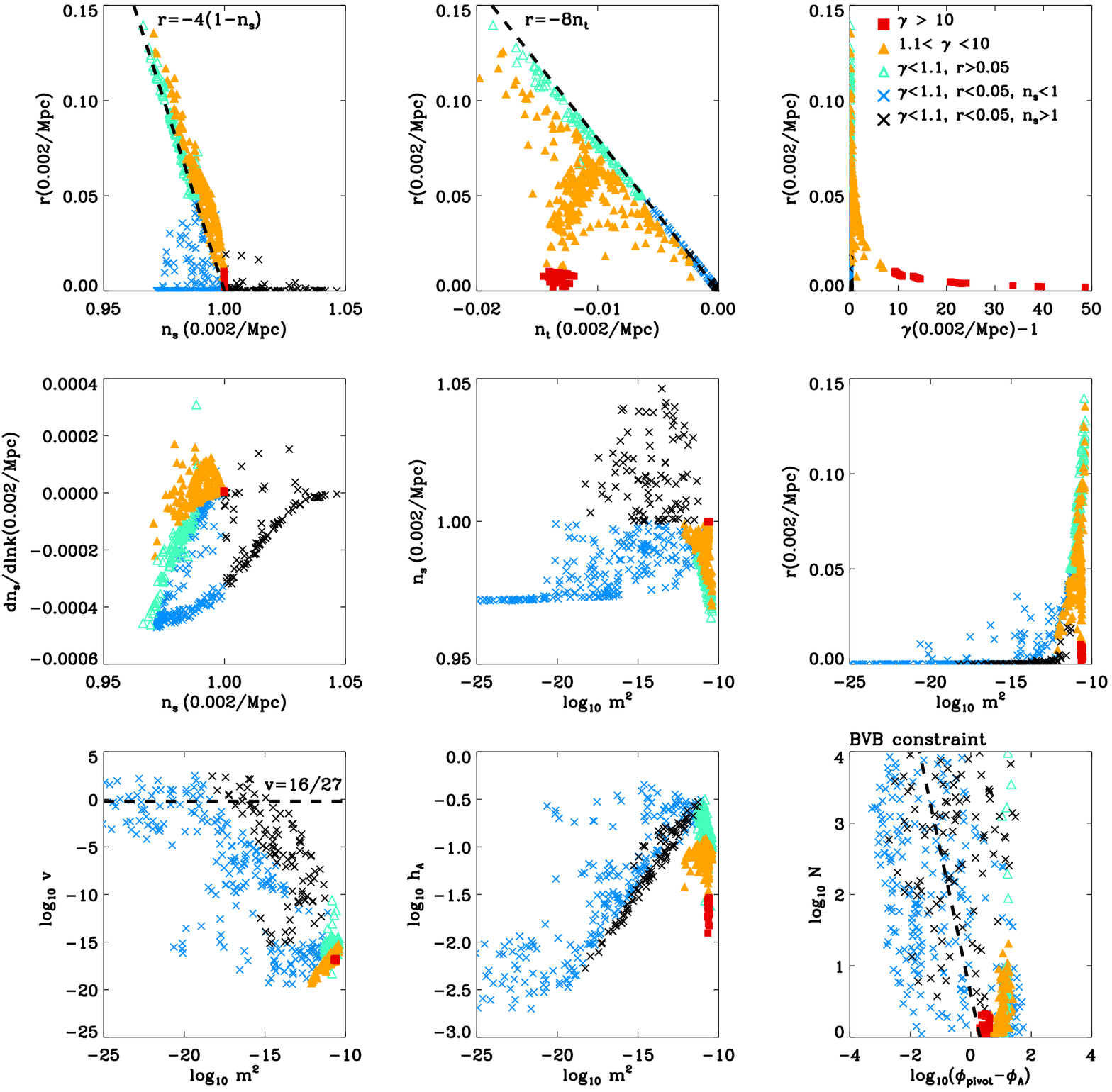} 
\caption{Taxonomy of the inflationary parameter space for the AdS warp geometry showing DBI inflationary models from a Monte Carlo simulation which satisfy the WMAP+SDSS+SN1a normalization constraint (\protect\ref{As_exp2}) at 95\% c.l.. The figure shows the wide variety of inflationary behavior arising from DBI inflation, including relativistic models ($1.1<\gamma<10$,filled yellow triangles and $\gamma>10$ full red squares), large tensor modes (open green triangles), and  blue and red-tilted slow-roll spectra (black and blue crosses, respectively). The taxonomy is presented in terms of relationships between the predicted spectrum observables ($n_{s}$, $r$, $dn_{s}/dlnk$) and key model parameters ($m^{2}$, $h_{A}$, $N_{A}$, $\gamma$,and $\phi_{pivot}$). Note that the bottom right figure shows models in comparison to the bulk volume bound imposed by \cite{Baumann:2006cd}. If the bound is imposed only slow-roll (low tensor, small running, non-relativistic models) with $\phi_{pivot}-\phi_{A}<\sqrt{4/N_{A}}$ are allowed. See the main text for further discussion. }
\label{adsmorph}
\end{center}
\end{figure}

\begin{figure}[t]
\begin{center}
\leavevmode
\includegraphics[height=0.7\textheight,angle=0]{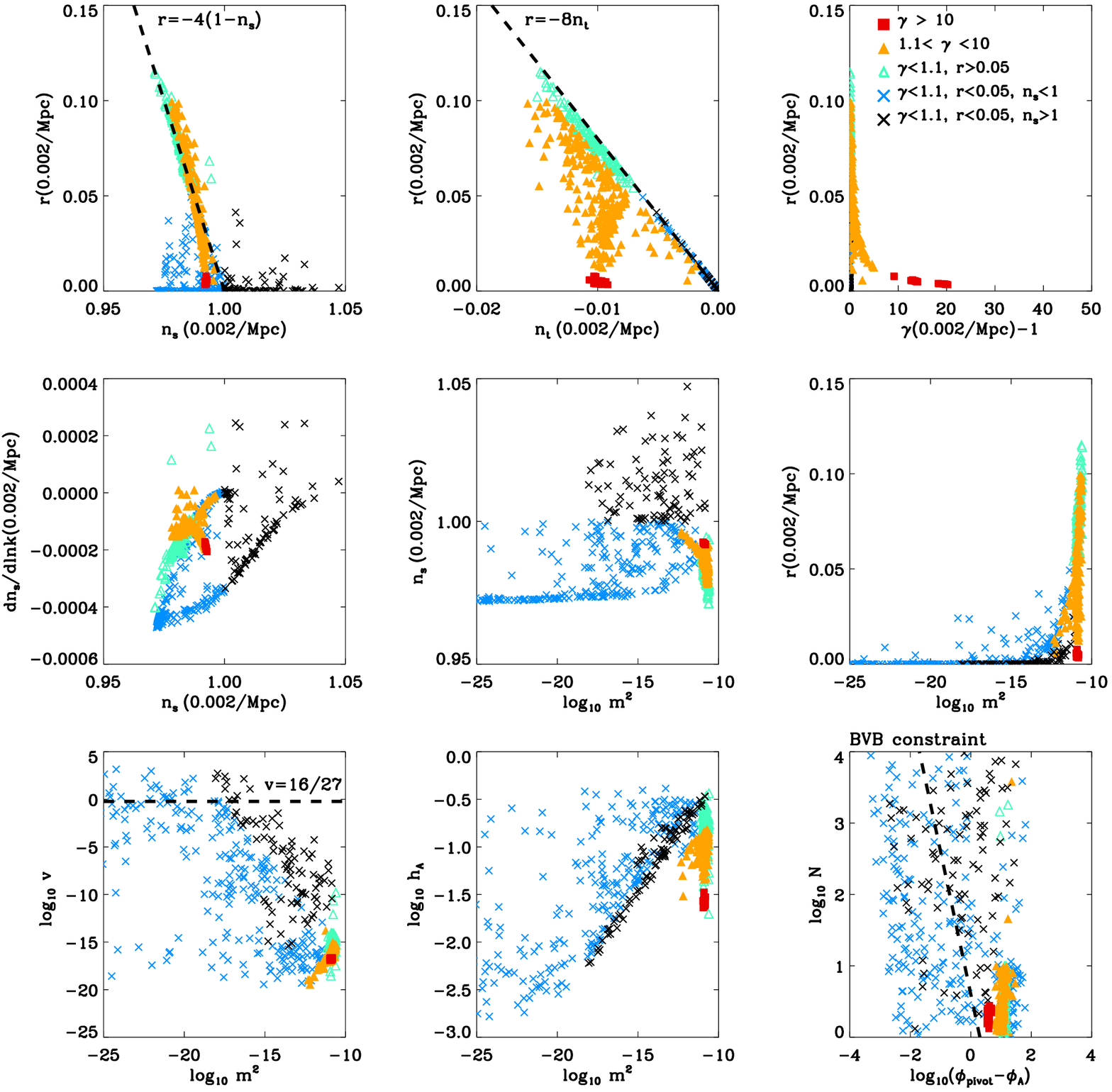}
\caption{As in Figure \protect \ref{adsmorph} but for the mass-gap warp geometry.}
\label{mgmorph}
\end{center}
\end{figure}

We compare the model parameters, for both the AdS and mass-gap geometries, to observational limits on the amplitude, spectral index, tensor-to-scalar ratio for the initial power spectra from the WMAP CMB data \cite{Spergel:2006hy,Hinshaw:2006ia,Jarosik:2006ib,Page:2006hz}, SDSS LRG matter power spectrum data\cite{Tegmark:2006az} and SNLS supernovae \cite{Astier:2005qq}.  We use the publicly available \texttt{cosmomc} code \cite{Lewis:2002ah} to perform a Monte Carlo Markov Chain analysis for a $\Lambda$CDM cosmological scenario, assuming flat priors on the following set of cosmological parameters: the physical baryon and cold dark matter densities, $\omega_b=\Omega_bh^2$ and $\omega_c=\Omega_ch^2$ (where $h$ is the reduced Hubble constant $h=H_{0}/100$), the ratio of the sound horizon to the angular diameter
distance at decoupling, $\theta_s$, and the optical depth to reionization, $\tau$, as well as power spectrum parameters, $n_{s}(k_{pivot})$,  $\log \ A_{s}(k_{pivot})$, and $r(k_{pivot})$.  As we discuss below, and in figures  \ref{adsmorph} and \ref{mgmorph}, in both the AdS and mass-gap scenarios, the running in the scalar spectral index is extremely small over observable scales, $|dn_{s}/d \ln k| < 4\times 10^{-3}$, and the vast majority models closely obey the consistency relation $n_{t}=-r/8$. We compare the Monte Carlo DBI models to compatible observational constraints obtained with $dn_s/d \ln k=0$ and both with the consistency relation, and with $r$ and $n_t$ allowed to be unrelated with $-0.02<n_t<0$. We consider purely adiabatic initial conditions and impose flat spatial curvature. The MCMC convergence diagnostics is done on seven chains though the Gelman and Rubin ``variance of chain mean''$/$``mean of chain variances'' $R$ statistic for each parameter.

We also impose the following two consistency constraints:
\begin{enumerate}
\item {\it Sufficient e-folds and all e-folds in the throat:} 
Although there may be inflation from other sources (other branes or other parts of the $D$3 trajectory) we require that sufficient e-folds to solve the usual horizon problem take place in a single throat. One may in principle loosen this requirement, but a more detailed knowledge of the geometry in the bulk space would be needed. The required number of e-folds depends on the reheating temperature $T_{RH}$ and inflation scale $H_I$, as outlined (\ref{min_efold}) in appendix \ref{app2}, roughly as $N_e\approx 68.6 +\ln (H_I \gamma / T_{RH})$. This equation comes simply from the requirement that $Ha$ at the beginning of inflation is at least as small as $Ha$ today, where $H$ is the Hubble parameter and $a$ is the universe's expansion factor. In other words, the largest scale we observe today ($k_H\sim k_{pivot}/10$) was once in causal contact. The expression is approximate because it assumes $H$ is constant during inflation. In the appendix we discuss how to implement the condition more precisely. The requirement that enough inflation occurs in the throat can then be expressed as $\phi_{H}<\phi_{UV}$, where $\phi_{UV}$ is the scale where the throat is glued to the bulk. We use the condition $h(\phi_{UV})=1$. For simplicity we assume 100\% efficient reheating, so that $T_{RH}$ is set by the warped brane tension. 

\item {\it A proportionally small throat:}
In string theory, the reduced Planck mass $M_{P}$ is related to the (warped) volume $V_{6}$ of the 6-dimensional bulk via $M_{P}^{2} = V_{6}/(2\pi)^{6} \pi g_{s}^{2}\alpha^{'4}$ So the position of the $D$3-brane cannot be physically larger than the largest dimension of the compactified bulk.
For consistency of the string theory model, the throat region should be small compared to the bulk of the compactified space. Other aspects of the model (like moduli stabilization details or physics in other throats), therefore, should be insensitive to the details of the inflationary throat. Assuming that the throat has the geometry of the KS throat or its orbifold, one can find the bound on how far the inflaton can be away from the bottom of the throat. This turns out to be a very tight constraint, as shown by Baumann and McAllister \cite{Baumann:2006cd}. One may put a minimal bound relating $M_p$ to the parameters $m_s$ and $N_A$ by calculating the warped throat volume. The condition can be usefully expressed as a bulk volume bound on the field range of the inflaton \cite{Baumann:2006cd},
\be
\frac{\phi_{pivot} - \phi_A}{M_P}<\sqrt{\frac{4}{N_A}}
\label{Liambound}
\ee
where $\phi_{pivot}$ is the value of $\phi$ corresponding to the physical pivot scale today, $k_{pivot}$, where the normalization to CMB data is applied. 

Together with (\ref{Lyth}) and even considering that $r$ may decrease rapidly during inflation, Ref.\cite{Baumann:2006cd} concludes that in DBI inflation with a quadratic potential, the tensor/scalar ratio would be unobservable for $N_A>10^6$ and would badly violate the current bound on non-Gaussianity for $N_A\gtrsim 40$. 
One may rewrite the Lyth bound (\ref{Lyth}) as
\ba
r_{\textrm{CMB}} &\lesssim& \frac{32}{N_A}\left(\frac{1}{\mathcal{N}_{\textrm{eff}}}\right)^2\\\nonumber
\mathcal{N}_{\textrm{eff}} &\equiv& \int_0^{\mathcal{N}_{\textrm{end}}}d\mathcal{N}\left(\frac{r}{r_{\textrm{CMB}}}\right)^{1/2}
\ea
Note that this definition has number of e-folds $\mathcal{N}$ and $\mathcal{N}_{\textrm{eff}}$ increasing with time. From (\ref{tsratio}),
\be
\frac{d\ln r}{d\mathcal{N}}=2(\epsilon+\kappa-\eta)
\ee
Then in the ultra-relativistic case, $\mathcal{N}_{\textrm{eff}}\sim\mathcal{O}(\epsilon^{-1})$. In the most optimistic observable case and with $\mathcal{N}_{\textrm{eff}}\sim1$,  $r_{\textrm{CMB}}\gtrsim10^{-4}$ requires $N_A<10^6$. One must then check whether the current bound on non-Gaussianity is violated. For our purpose, this constraint in the form of the bulk volume bound (\ref{Liambound}) turns out to be most useful. Note that this bound may be relaxed in some variations of the model that we discuss later.
\end{enumerate}

We impose the first condition directly in the code, but implement the last as a consistency cut. Since other geometries (e.g., a squashed throat) or other choices for the inflaton may have a different dependence on $N_A$, it is interesting to examine the range of possible observables before and after imposing this constraint.  

\begin{figure}[t]
\begin{center}
\leavevmode
\includegraphics[width=0.7\textwidth, angle=0]{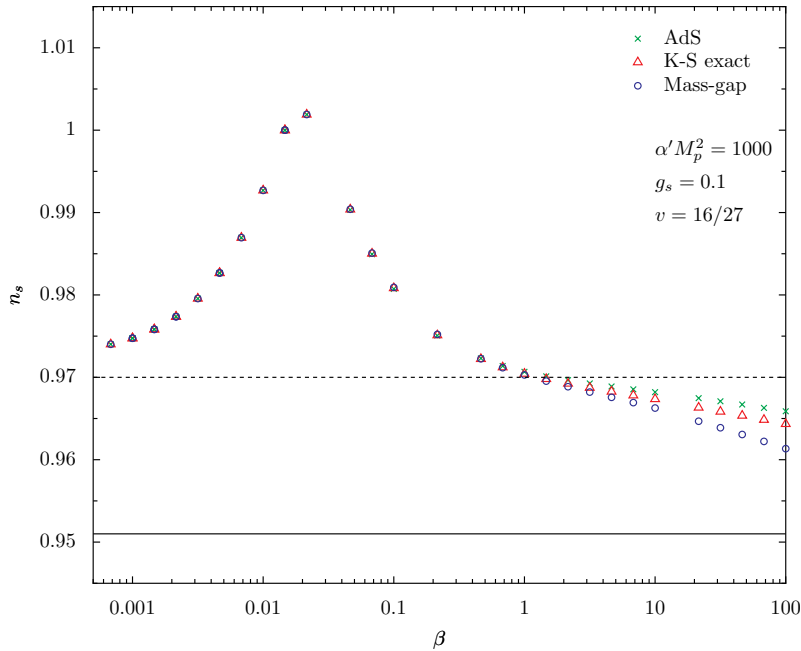}
\caption{Comparison of $n_s$ in different warped geometries.}
\label{fig_AdSmg}  
\end{center}
\end{figure}

In Figure \ref{fig_AdSmg} we demonstrate the sensitivity of $n_s(k_{pivot})$ to different warped geometries, holding all parameters, except $\beta$, at typical values and varying $\beta$ from slow-roll, $\beta\ll1$, to ultra-relativistic, $\beta\gg1$, regimes. We compare the AdS warp factor, the mass-gap warp factor and the exact Klebanov-Strassler warp factor.  We see that in the slow-roll regime, since the warp factor does not come into play, different warped geometries give the same prediction on $n_s$. When $\beta > 1$ the model starts to deviate from slow-roll and the warp factor comes into play in the equation of motion, yielding different predictions on $n_s$. Typically, we find that the mass-gap geometry gives a more red-tilted spectrum than the AdS geometry and the exact KS is in between the two.

\begin{figure}[!t]
\begin{center}
\leavevmode
\includegraphics[height=0.7\textheight,angle=0]{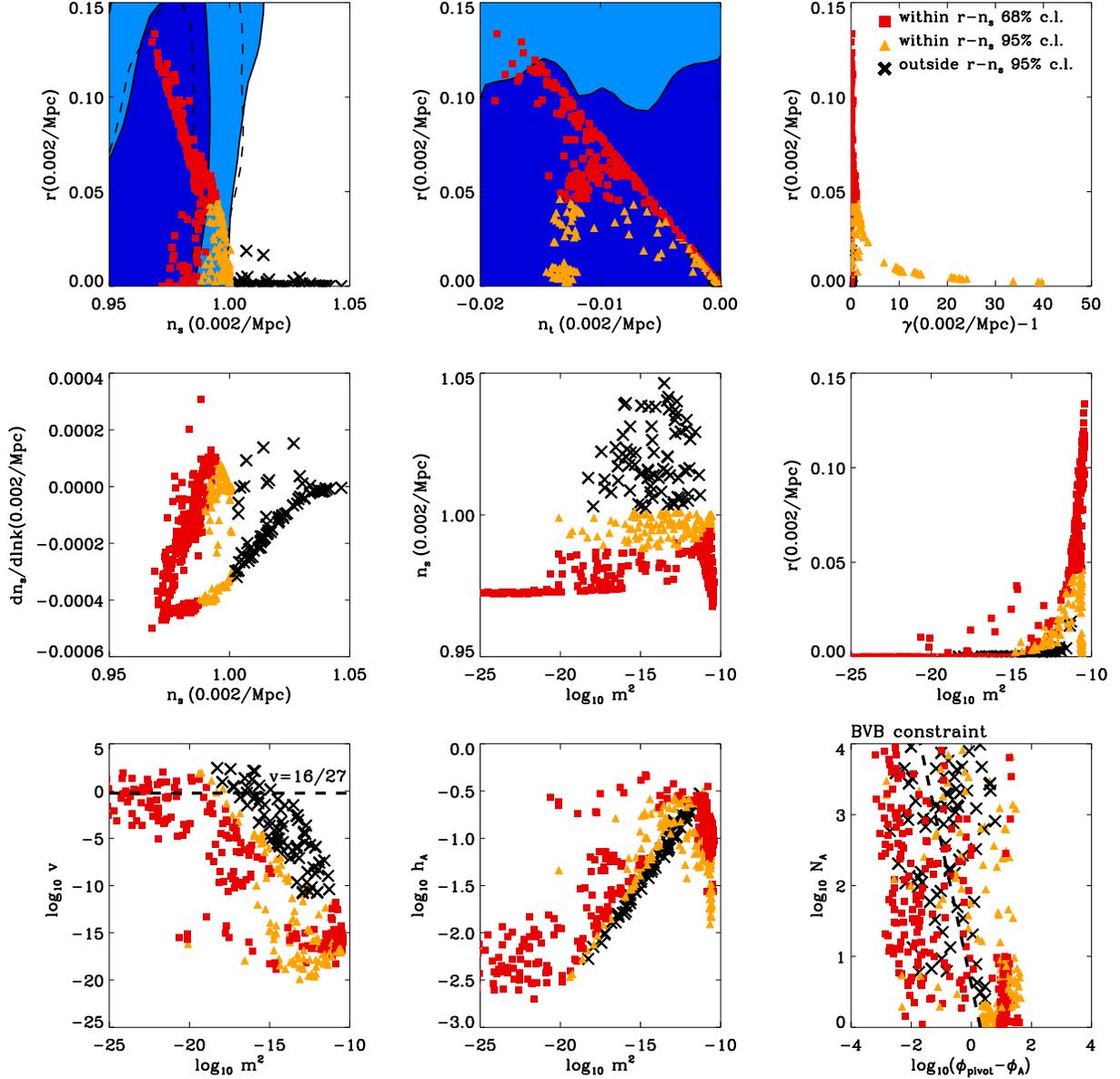} 
\caption{Inflation parameter space for the AdS geometry in comparison to cosmological observational constraints from 3-year WMAP data in combination with SDSS LRG power spectrum data and SNLS SN1a data. The top left figure ($n_{s}$ vs $r$ constraints) and top center figure ($n_t$ vs $r$ constraints) of each set show the observational constraints without imposing the $n_t=-r/8$ consistency relation, allowing $-0.02<n_t<0$, at the 68\% (dark blue) and 95\% (light blue) confidence level. Overlaid are DBI inflationary models from a Monte Carlo simulation which satisfy the WMAP normalization constraint(\protect\ref{As_exp2}) at 95\% c.l. and are in agreement with the $n_s - r$ WMAP+SDSS LRG+SN1a constraint within 68\% (red square), 95\% (yellow triangle) and outside 95\% (black cross) confidence limits. The constraints with the consistency relation imposed (dashed contours in the top left plot) yield a very similar splitting of the DBI models by confidence level.}
\label{adsfull}
\end{center}
\end{figure}

\begin{figure}[!t]
\begin{center}
\leavevmode
\includegraphics[height=0.7\textheight,angle=0]{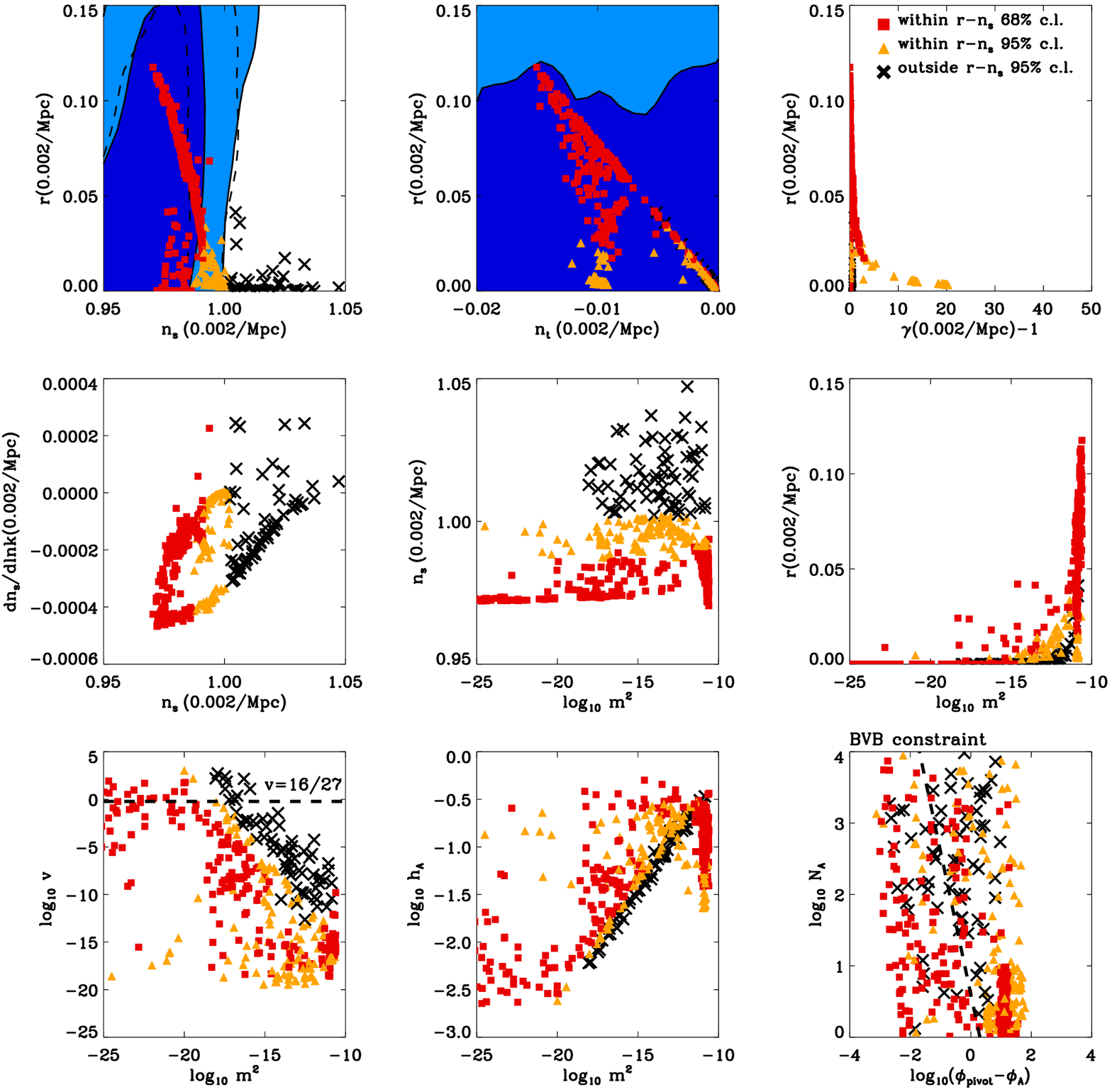}
\caption{As in Figure \protect \ref{adsfull} but for the mass-gap warp geometry.}
\label{mgfull}
\end{center}
\end{figure}

In Figures \ref{adsmorph} and \ref{mgmorph} we demonstrate the wide variety of observational properties arising in DBI inflation. In addition to slow-roll behavior with red tilted, $n_{s}<1$, and blue-tilted, $n_{s}>1$, spectra, DBI inflation can also give rise to relativistic behavior $\gamma>1$, and intermediate models with large tensors, small running and a strong running-spectral tilt relation.  The figures show key 2D parameter spaces for both observables and model parameters with Monte Carlo generated models which are consistent with the WMAP+SDSS+SN1a normalization (\ref{As_exp2}) at the 95\% confidence level (c.l.). The red squares are within 1$\sigma$ in the upper-left  $r-n_{s}$ panel,
the orange-yellow triangles are within 2$\sigma$, while the black crosses are outside 2$\sigma$.   
The nine panels from bottom left are:
\begin{enumerate}
\item $\log_{10}m^{2}-\log_{10}v$ panel. Low masses, $m^{2}\lesssim 10^{-12}M_p^2$, produce slow-roll behavior $\gamma\sim 1$ with small tensors, $r<0.05$, and obeying the usual consistency relation $r=-8n_{t}$. Imposing  a flat prior on $\log_{10}m^{2}$ and $\log_{10}v$, the majority of slow-roll models have a red-tilted spectrum. The normalization and e-folding constraints allow the orbifolding parameter, $v$, to extend over a large range of values, $\sim 10^{25}$ orders of magnitude. The intermediate and relativistic regimes are constrained in a small section of this space with very small orbifold factors $v\lesssim 10^{-17}$. As such, only the slow-roll regime is consistent with the Klebanov-Strassler Throat for which $v=16/27$.
\item $\log_{10}m^{2}-\log_{10}h_{A}$ panel. Two distinct regions, the slow-roll regime for $m^{2}\lesssim10^{-12}M_p^2$ and a relativistic region with $m^{2}\gtrsim 10^{-11}M_p^2$ are carved out. The spread of $h_{A}$ at small $m^{2}$ indicates the insensitivity of data to the warp factor of the throat.  The requirement that there are sufficient e-foldings to include the physical scale $k_{pivot}$ gives the lower bound on $h_{A}$ for a given mass in the slow roll region, with the blue tilted slow-roll models lying close to this boundary. In the intermediate and relativistic regimes the normalization constraint translates into a thin region of allowed models. The relativistic models require a small warp factor at the bottom of the throat, $h_{A}\sim 10^{-1.5}$.
\item $\log_{10}(\phi_{pivot}-\phi_{A})-\log_{10}N_{A}$ panel. Conveniently demonstrates the bulk volume bound (\ref{Liambound}).  The relativistic models all have $(\phi_{pivot}-\phi_{A}) / M_p > \sqrt{4/N_{A}}$, as such, imposing the bound rules out all but the slow-roll regime. 
\item $dn_{s}/dlnk - n_{s}$ panel. For both the AdS and mass-gap geometries the bulk of the models satisfying the e-fold and normalization bounds have a very small negative or near zero running in the spectral tilt, $-0.001<dn_{s}/dlnk(k_{pivot})<0$. The relativistic models, in particular, have extremely small running and $n_{s}\approx 1$ as discussed in regards to (\ref{nsm1}). 
\item $\log_{10}m^{2}-n_{s}$ panel. For $g_{s}=0.1$ and $\alpha' M_p^2=1000$ the slow-roll regime tends towards a common spectral tilt $n_{s}\approx 0.972$ as $m^{2}\rightarrow 0$, which reproduces the spectral index in the KKLMMT scenario \cite{Kachru:2003sx}. 
\item $\log_{10}m^{2}-r$ panel. Models with significant tensor contributions with $r>0.05$ are purely generated in the intermediate regime with $m^{2}\sim 10^{-11}M_p^2$.
\item $n_{s}-r$ panel. Models with a noticeable deviation from the Harrison-Zeldovich spectrum ($n_s=1$ and $r=0$) are possible with $0.96\lesssim n_{s}\lesssim 1.05$, including models with significant tensor contributions $0<r\lesssim 0.15$. (\ref{nsm1}) and (\ref{tsratio}) lead to the relation $4(1-n_s)/\gamma\sim r$. The AdS Monte Carlo models are able to have slightly higher tensors than the mass-gap geometry. The mass-gap, however, allows an interesting branch of models in the intermediate regime with large tensors and distinct $n_s$ vs. $r$ relation.
\item $n_t-r$ panel. Slow-roll and many intermediate tensor mode models are in good agreement with the $r=-8n_t$ consistency relation. Significant deviations from this relation are seen in the relativistic models where the models have a significant tensor spectral tilt, but comparatively much smaller tensor amplitude given by $r\sim n_{t}/8\gamma$ in (\ref{ntgamma}).
\item $\gamma - r$ panel. There is a the clear division of behavior between the large mass models, with intermediate models possessing large tensor modes but a comparatively slowly rolling inflaton $\gamma\sim 1$, and relativistic scenarios with $\gamma> 1$ but vanishing tensor contributions in the limit $\gamma\gg1$.  For large $\gamma$, the Monte Carlo models are subject to additional constraints from CMB non-Gaussianity measurements, $\gamma<31$.
\end{enumerate}

In Figures \ref{adsfull} and \ref{mgfull} we compare the observational constraints from 3-year WMAP CMB + SDSS +HST GOODs data with the Monte Carlo results for DBI inflation for the AdS and mass-gap warp factor respectively without imposing a constraint on $\phi_{pivot}$. The panels show the regions within 1$\sigma$ (dark blue) and  2$\sigma$ (pale blue) allowed by the data. The DBI models are marked red (yellow/orange) if they lie within 1$\sigma$ (2$\sigma$) of the data in the key $n_s-r$ space. Black points are outside the 2$\sigma$ region. It is clear that the data is constraining the parameter space of the model in a meaningful way. 

One can see that in both cases the inflationary parameter space \{$N_A, v, h_{A},m^{2}$\} yields models wholly consistent with observations. For $g_{s}=0.1$ and $\ap M_p^2=1000$, the slow-roll, low mass, regime tends towards a common spectral index $n_{s}(k_{pivot})\approx 0.972$, consistent with observations at the 1$\sigma$ level. An intermediate regime, for masses $m^{2} \sim 10^{-11}M_p^2$ and  $N_{A}\sim 1$, with a significant tensor contribution, is also consistent at 1$\sigma$ level. The relativistic ($\gamma \gg1$) regime is consistent with data at the 95\% c.l. For the AdS warp factor a handfull of Monte Carlo models, in agreement with the CMB and galaxy spectra at the 95\% level, exceed the non-Gaussianity constraint from WMAP CMB on relativistic models, $\gamma < 31$.

\begin{figure}[!t]
\begin{center}
\leavevmode
\includegraphics[height=0.7\textheight,angle=0]{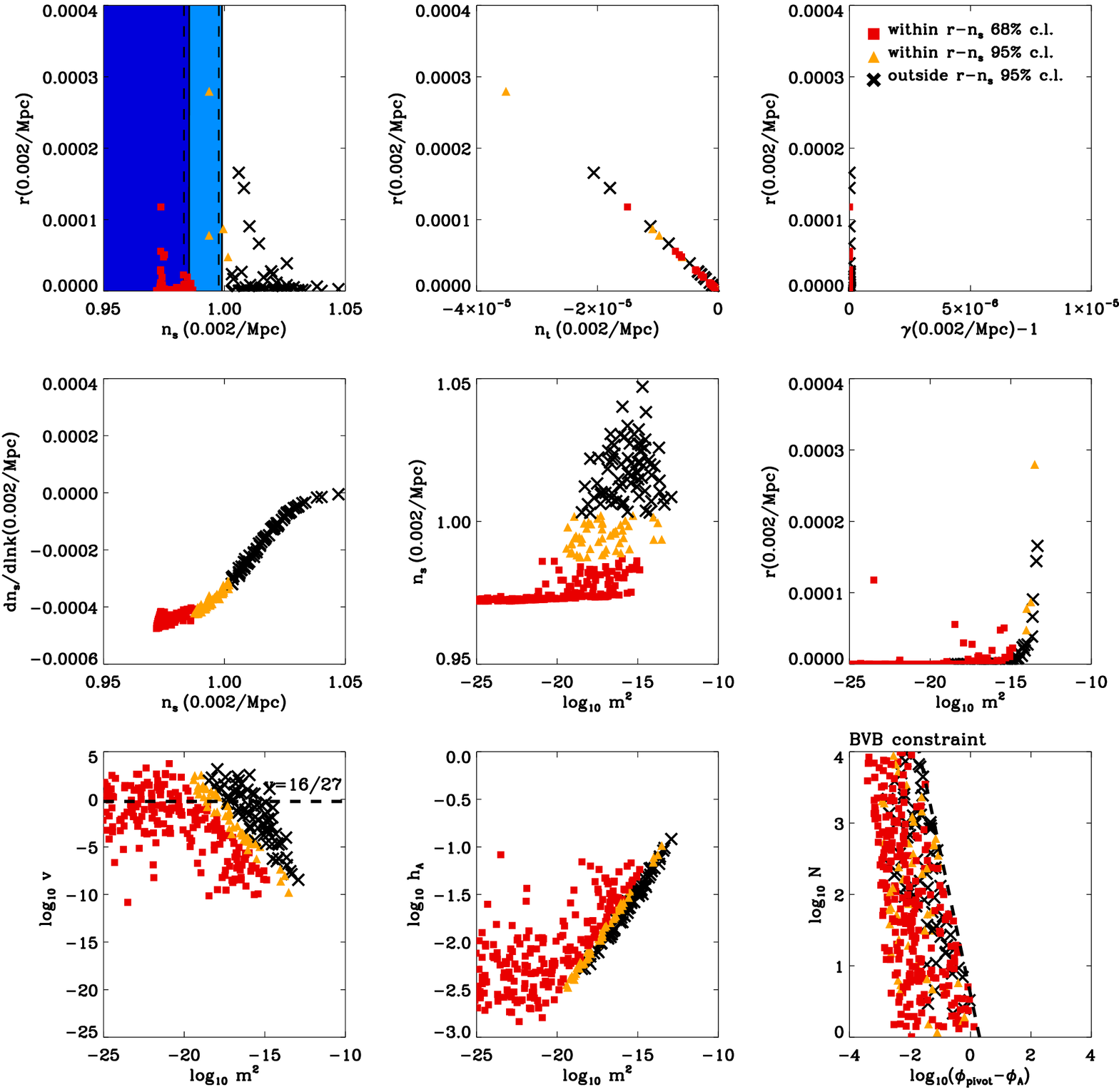} 

\caption{As in Figure \protect \ref{adsfull} but imposing the bulk volume bound \cite{Baumann:2006cd} of $\phi_{pivot}-\phi_{A}< \sqrt{4/N_{A}}$ for  the AdS warp factor. Note that the tensor ratio and tensor tilt scales are reduced in comparison to Figure \protect \ref{adsfull}. The rescaled plots show that the, much more restricted, set of models are all in the slow-roll regime with $r<0.003$ and $\gamma -1< 10^{-7}$ and satisfy the $n_{t}=-r/8$ consistency relation. The intermediate regime, while in strong agreement with observations, is incompatible with the bound.} 
\label{adscrop}
\end{center}
\end{figure}

\begin{figure}[!t]
\begin{center}
\leavevmode
\includegraphics[height=0.7\textheight, angle=0]{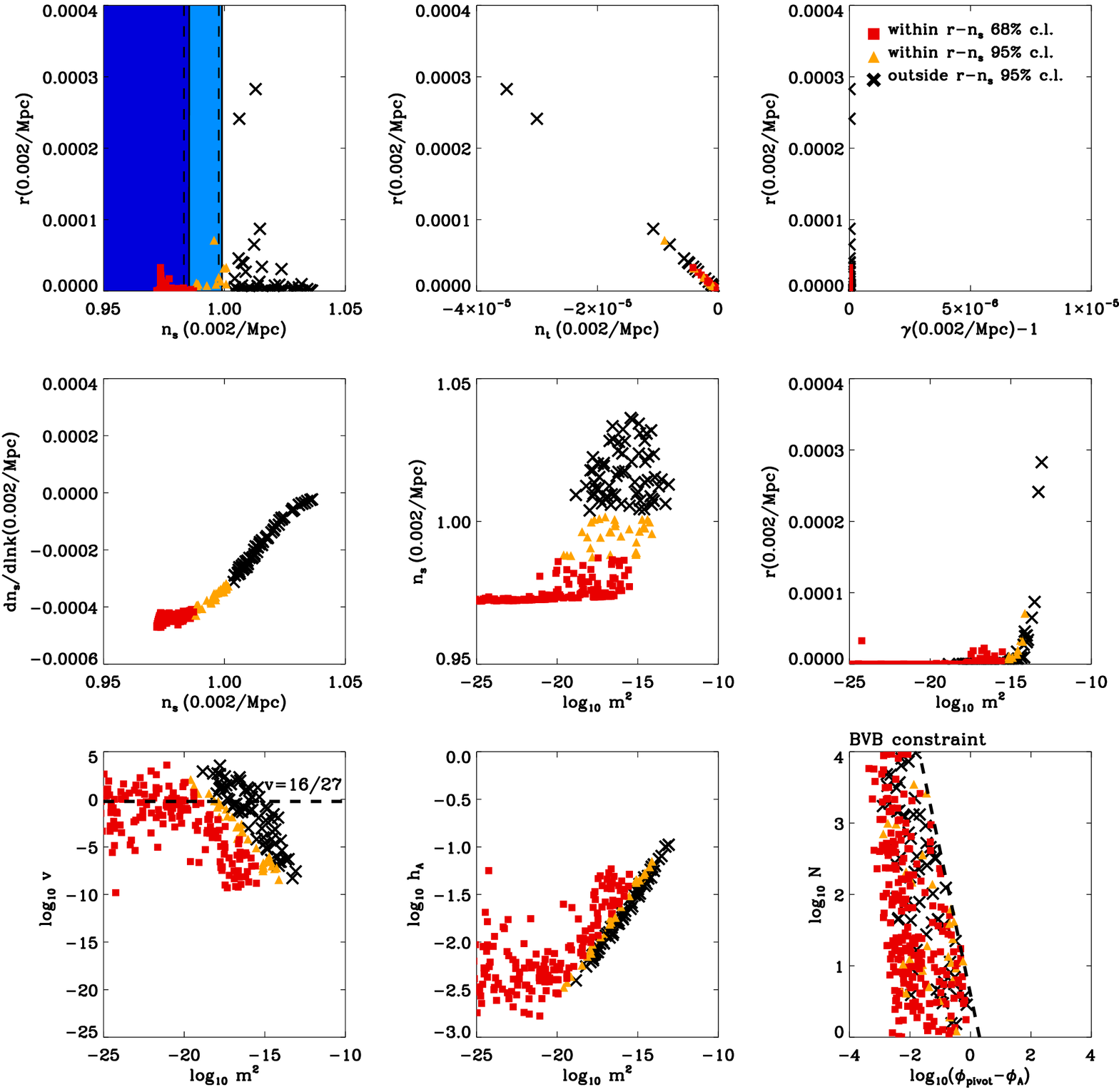}
\caption{As in Figure \protect \ref{adscrop} but imposing the bulk volume bound for the mass-gap geometry.  Only slow-roll models remain consistent with the data  at the $95\%$ confidence level with $r<0.001$ and $\gamma -1< 10^{-7}$ and satisfy the $n_{t}=-r/8$ consistency relation.} 
\label{mgcrop}
\end{center}
\end{figure}

\begin{figure}[t]
\begin{center}
\leavevmode
\includegraphics[width=1.\textwidth, angle=0]{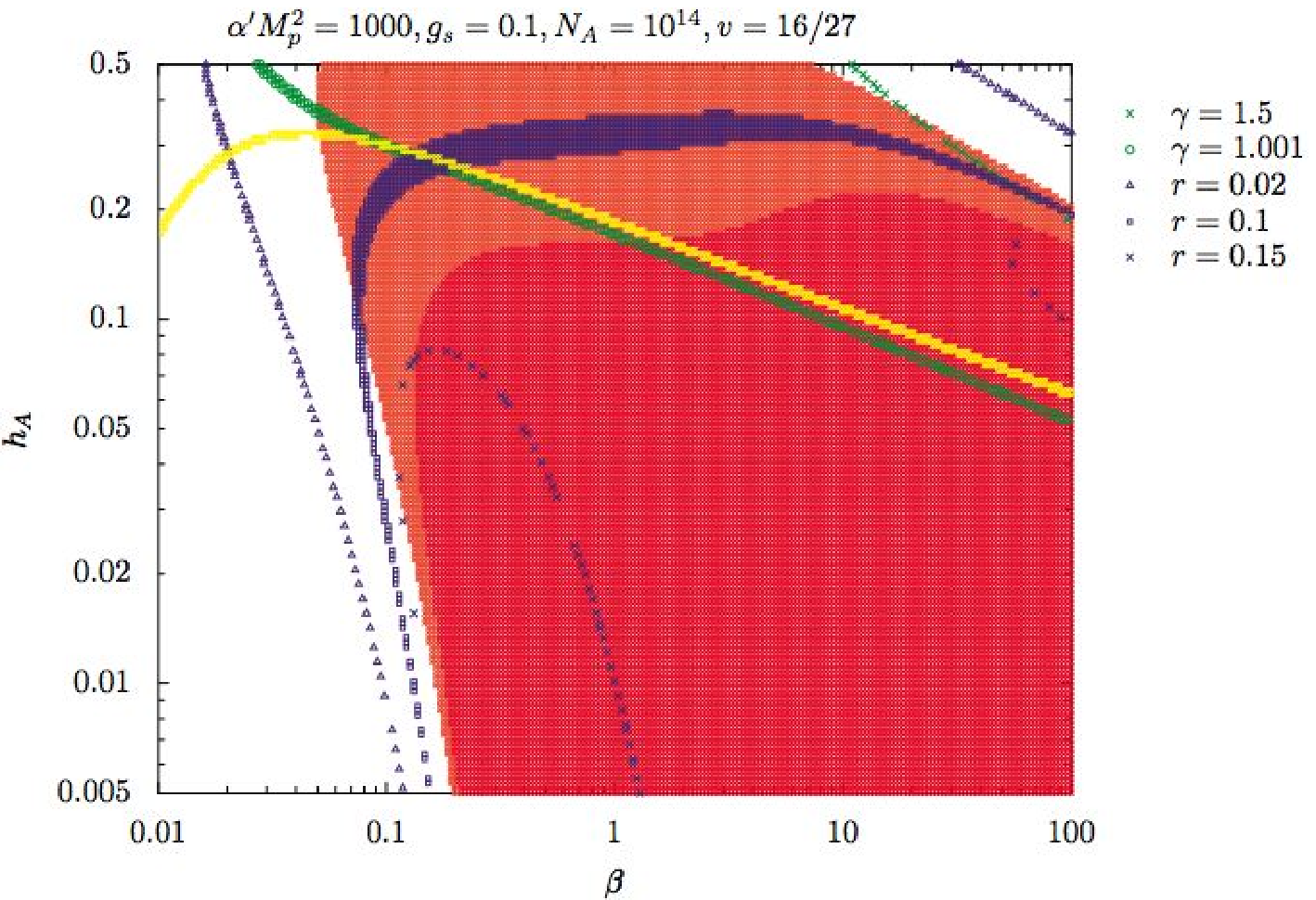} 
\caption{With $N_A=10^{14}$, the figure shows the $\beta - h_A$ plane of parameter space for the AdS warp factor without orbifolding. The red region fits $n_s$ within $2\sigma$, and the yellow band is the constraint from COBE normalization. We find that we need $N_A$ roughly $10^{14}$ to fit the WMAP data. The blue lines are contours of $r = 0.01, 0.1, 0.15$. The model yields significantly large tensor to scalar ratio. The green lines are contours of $\gamma = 1.001, 1.5$. All the models within the red region deviate from the slow-roll scenario, but $\gamma$ at the pivot scale is not large enough to be in the ultra-relativistic regime. They all belong to the intermediate region. The tightest constraint here is actually the bulk volume bound (\protect \ref{Liambound}), with $N_A \sim 10^{14}$ : no points on this figure survive the bound.} 
\label{fig_largeN}
\end{center}
\end{figure}

\begin{figure}[b]
\begin{center}
\leavevmode
\includegraphics[width=1.\textwidth, angle=0]{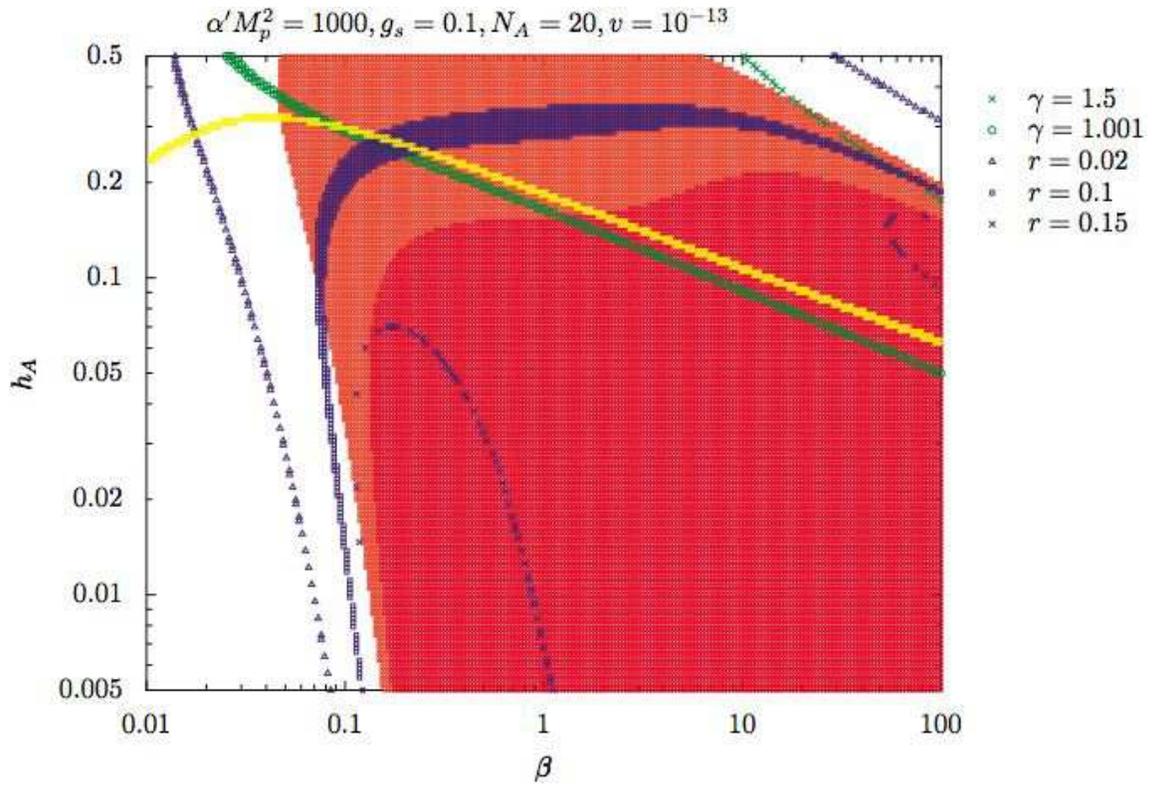}
\caption{A similar graph as Figure \protect \ref{fig_largeN} except that the volume of $X_5$ is decreased to $v = 10^{-13}$. We find models that fit WMAP data, and they come with much smaller $N_A$ values than the case with $v \sim 1$. The graph shows a plane with $N_A = 20$. This figure is still in the intermediate region. However, even if $N_A$ is decreased significantly, no model survives the bulk volume bound (\protect \ref{Liambound}).} 
\label{fig_smallN}
\end{center}
\end{figure}

\begin{figure}[t]
\begin{center}
\leavevmode
\includegraphics[width=1.\textwidth, angle=0]{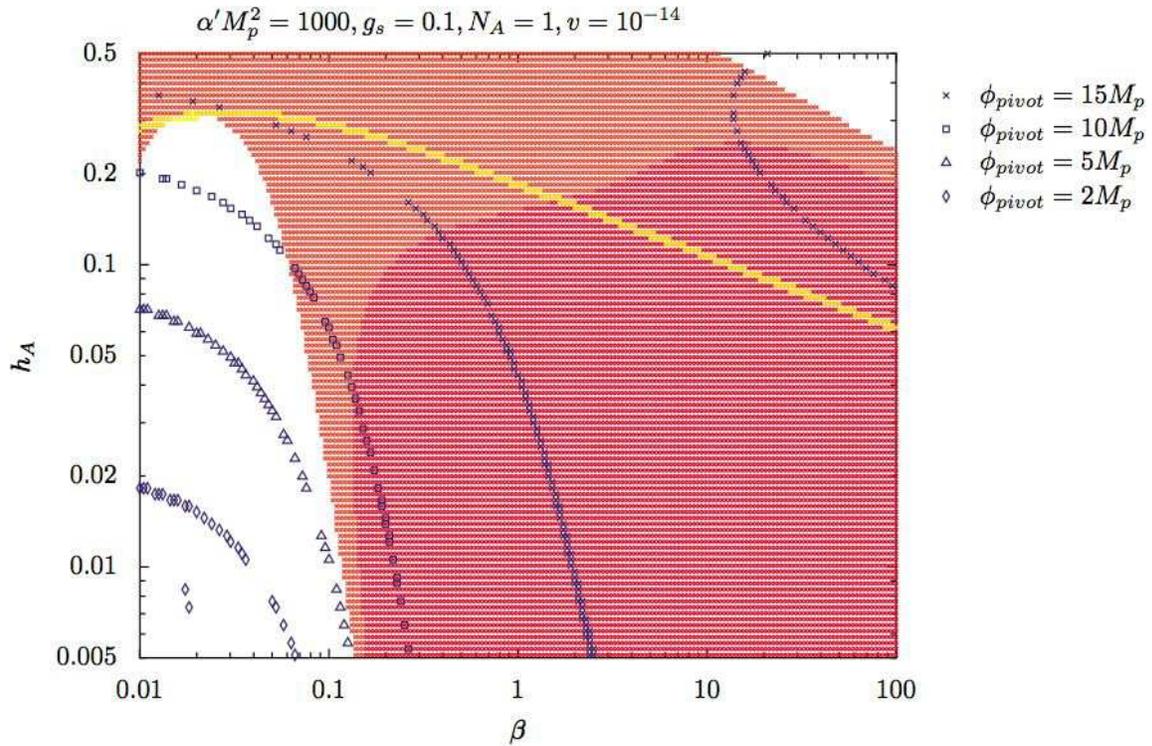}
\caption{We now take the limit $N_A = 1$ and see if there are any models that fit the data and survive the bulk volume bound (\protect \ref{Liambound}) at the same time. The blue lines are contours for different $\phi_{pivot}$ values. (Blue-tilt at small $\phi_{pivot}$.) In the $2\sigma$ region of the $n_s$ constraint, i.e. the red region, we see that $\phi_{pivot} > 10$, badly violating the bound on field range. The diamond points do show some points that survive the bound but they do not fit $n_s$ well. The figure shows that the value of $\phi_{pivot}$ decreases in the upper-right and lower-left corner, so for $\beta > 10^2$ and $\beta < 10^{-2}$, the chance to get small $\phi_{pivot}$ is much larger, and these two regions are the slow-roll region and the DBI region.}
\label{fig_pivot}  
\end{center}
\end{figure}

In Figures \ref{adscrop} and \ref{mgcrop} we show the allowed Monte Carlo models after additionally requiring that they also satisfy the bulk volume bound (\ref{Liambound}). This bound effectively constrains the tensor to scalar ratio, $r(k_{pivot})<4\times 10^{-3}$. In this case the slow-roll regime is allowed but the intermediate and relativistic regime do not satisfy the cut. In Figures \ref{fig_largeN} - \ref{fig_pivot} we investigate the intermediate regime more fully, for a variety of specific parameter values, and find that the bulk volume constraint disallows any of the observationally consistent models to survive.  

\begin{figure}[t]
\begin{center}
\leavevmode
\includegraphics[width=0.7\textwidth, angle=0]{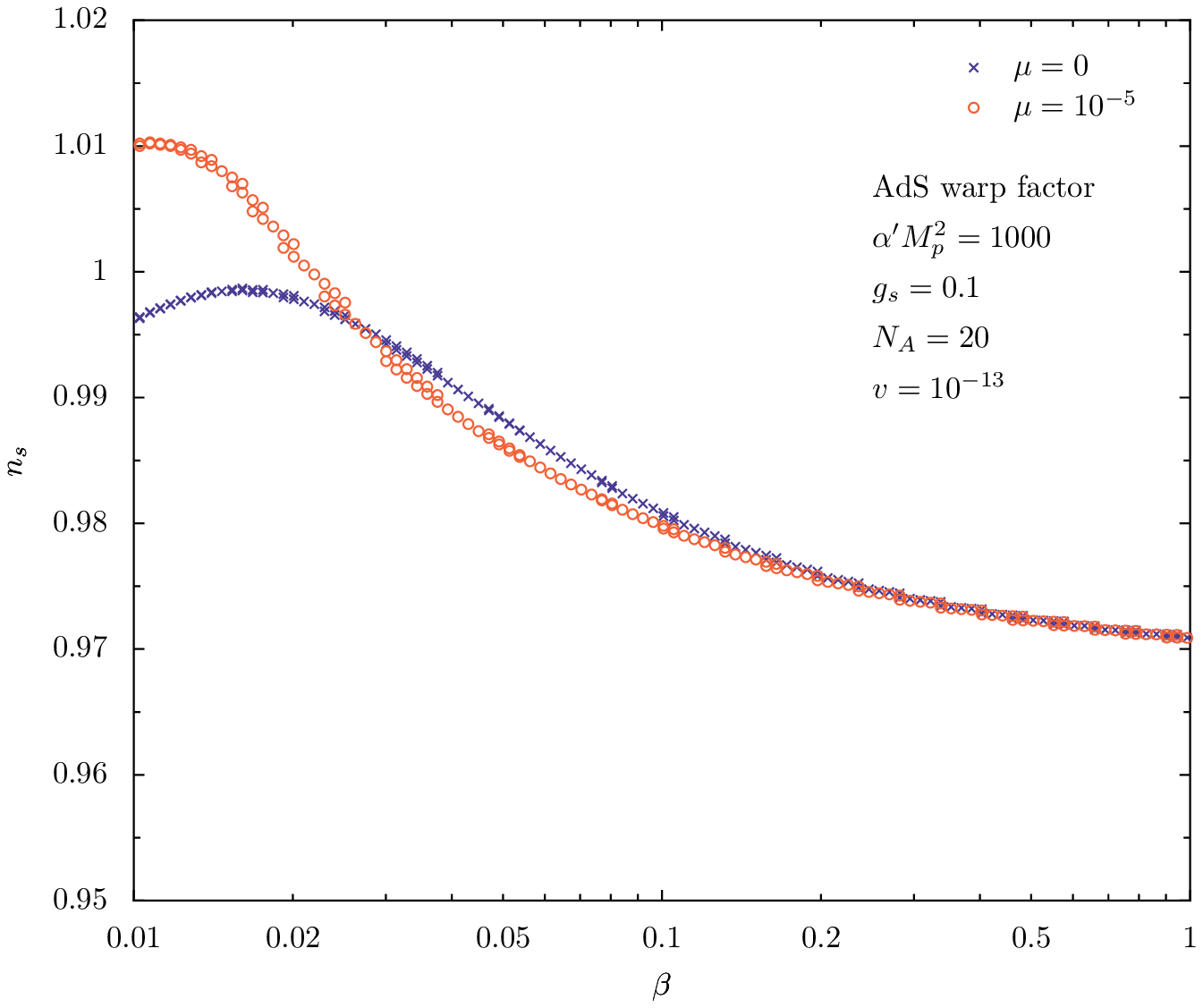}
\includegraphics[width=0.7\textwidth, angle=0]{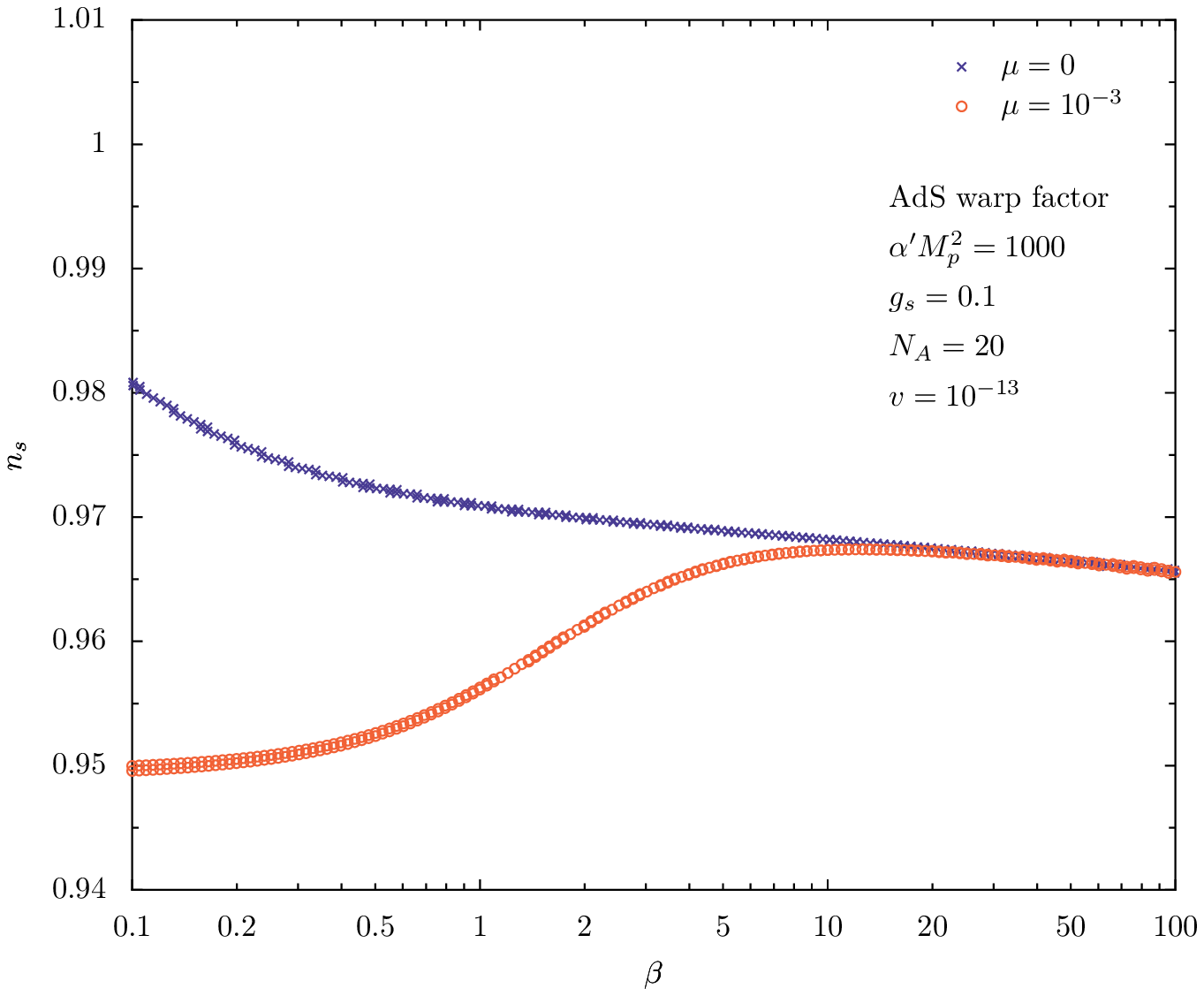}
\caption{These two figures show the prediction of $n_s$ if the inflaton potential is perturbed by a $\mu \phi^4$ term. In the slow-roll regime ($\beta < 0.1$), a $\mu \phi^4$ term will drive the $n_s$ to the blue end and is not favored by data. But in the intermediate regime ($\beta > 0.1$), adding a $\mu \phi^4$ term in the potential drives $n_s$ to the red end and improves the model's agreement with data. }
\label{figlambda} 
\end{center}
\end{figure}

For the intermediate region, it is worth noting that the initial $\phi$ takes a rather large value. In this case, we expect that the truncation, keeping only the $m^{2}\phi^{2}$ term in the inflaton potential, may not be a good approximation. Let us, therefore, consider the impact of introducing a $\phi^{4}$ term to the inflaton potential (\ref{inflatonpot}) :
\be
V(\phi) \rightarrow V(\phi) + \mu V_{0} \left(\frac{\phi}{M_{p}}\right)^{4}
\ee
We see in Figure \ref{figlambda} that turning on a $\phi^{4}$ term tends to blue (red) tilt the power spectrum for small (intermediate) values of $m^{2}$. 

\begin{figure}[t]
\begin{center}
\leavevmode
\includegraphics[width=0.7\textwidth, angle=0]{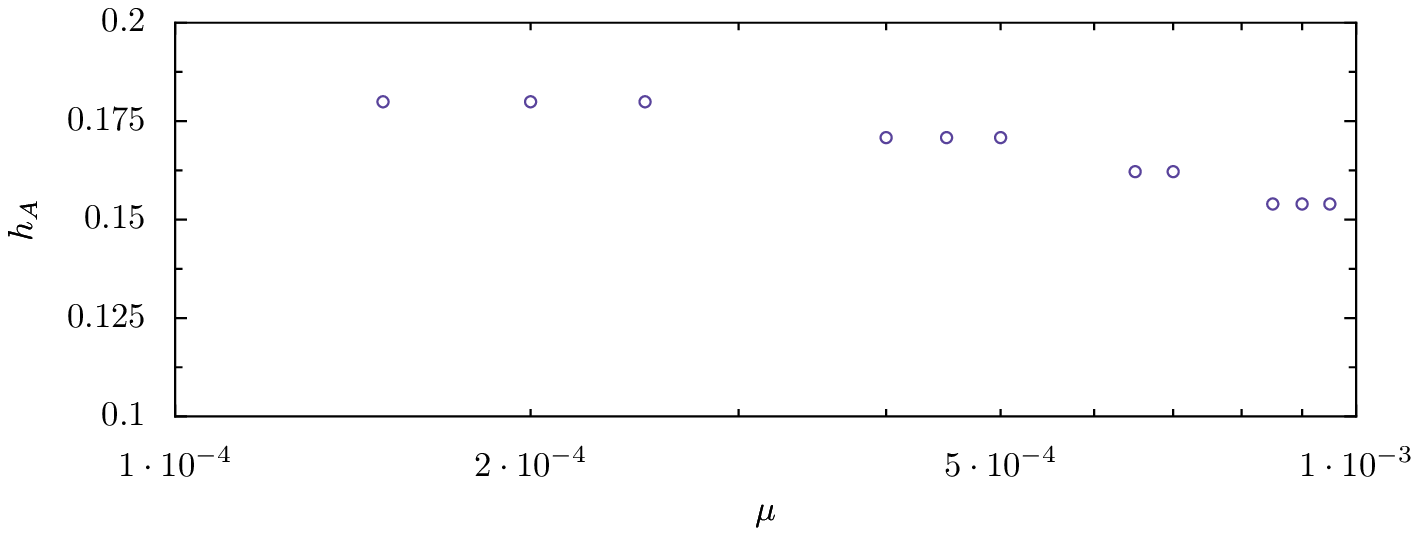}
\includegraphics[width=0.7\textwidth, angle=0]{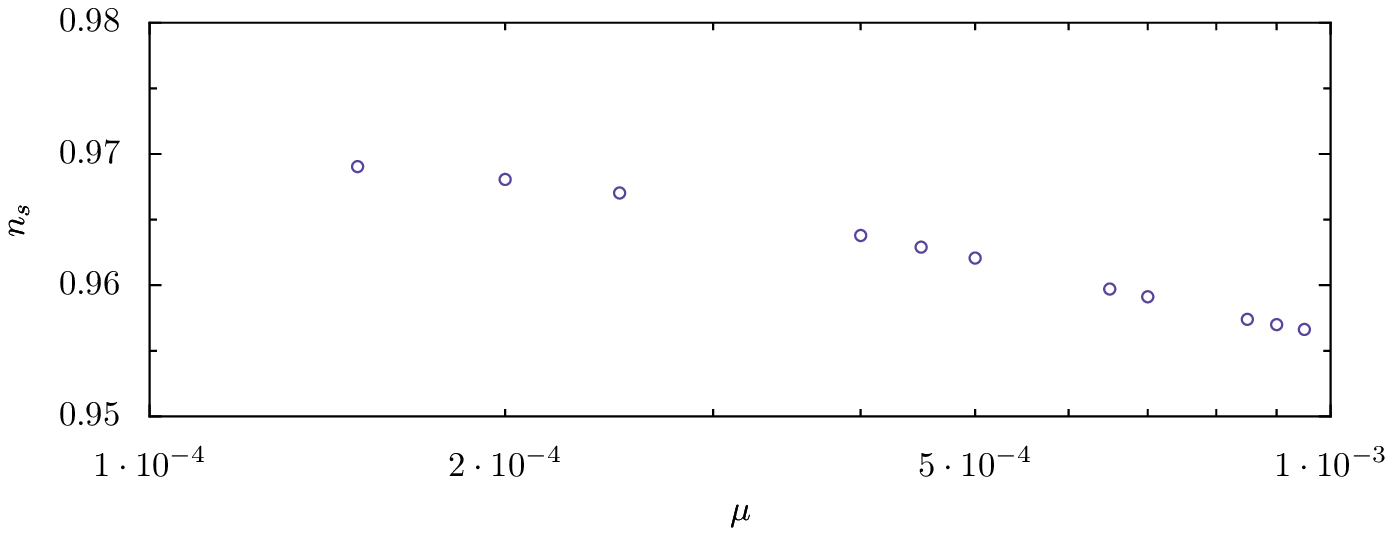}
\includegraphics[width=0.7\textwidth, angle=0]{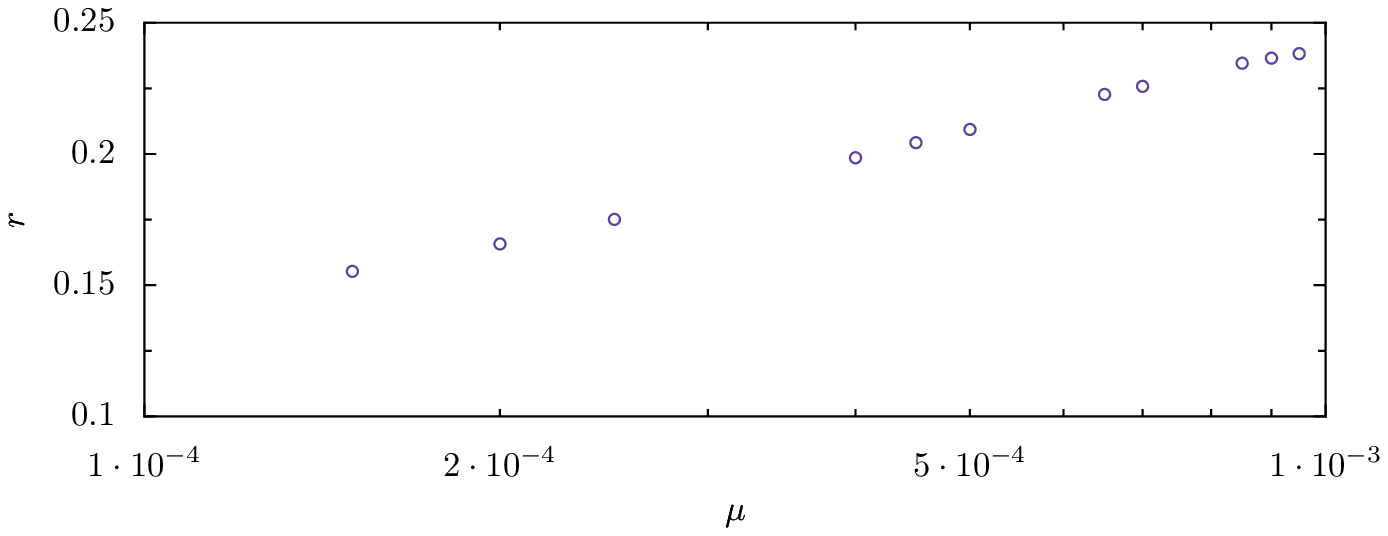}
\includegraphics[width=0.7\textwidth, angle=0]{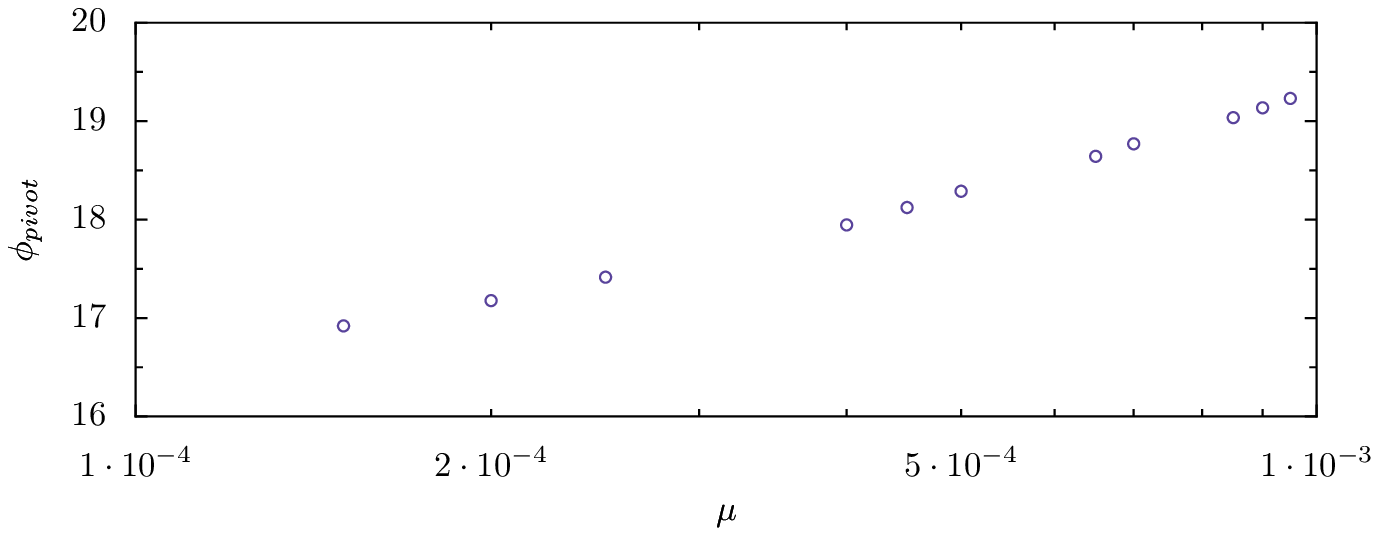}
\caption{For $\beta = 1$, these figures show the dependence of $\{h_A, n_s, r, \phi_{pivot} \}$ on the $\mu\phi^{4}$ term. The other parameters are fixed to be $N_A = 20$, $v = 10^{-13}$, $g_s = 0.1$, $\alpha' M_p^2 = 1000$.}
\label{param_mu}
\end{center}
\end{figure}

In Figure \ref{param_mu}, we show how the parameters and observables change with different values of $\mu$. We find that as $\mu$ increases, $h_A$ has to decrease in order to fit COBE normalization, meanwhile, $n_s$ is driven to the red end and $r$ increases.  We also show in the figure that including the $\phi^4$ term does not help to reduce the value of $\phi_{pivot}$ but makes it even larger, consistent with the bound (\ref{Lyth}) as $r$ increases with larger $\mu$.

\section{Possible scenarios with a large tensor perturbative mode}\label{largetensor}

The intermediate region is severely constrained by the limited size of the bulk. If not for this bulk volume bound (\ref{Liambound}), we see that the tensor mode can be substantial. However, we would like to point out that there are ways in which future realistic models may in fact explore a wider parameter range. Let us discuss a couple of possibilities where the bulk volume bound may be relaxed enough to allow a large tensor mode. 

To see how to bypass/relax the bulk volume bound, it is useful to first recap the reason why it is so constraining. In a typical chaotic inflationary scenario, the inflaton field value can be much larger than the Planck mass $M_{p}$ and the 2 slow-roll parameters are comparable,
$\epsilon_{SR} \sim \eta_{SR}$. Since $r =16 \epsilon_{SR}$, we see why the tensor mode can be large (i.e., $r$ is a few percent or larger). In brane inflation, the inflaton, which is the brane position, cannot be bigger than the bulk size, that is $\phi < M_{p}$. To get enough inflation, it must move slowly. The Lyth bound (\ref{Lyth}) then tells us that $r$ must be small if we want to get enough e-folds from moving down a throat. This also explains why $\epsilon_{SR} \ll \eta_{SR}$ in this small field model. 

One way to satisfy the bulk volume bound is if we need only a few e-folds coming from the motion of $\phi$ down the throat; a substantial number of e-folds are coming after $\phi$ has moved down the throat. This leads us to consider the following scenario.

$\bullet$ {\it Multi-brane-multi-throat scenario} Perhaps the simplest way to generalize the single-throat scenario is to loosen the requirement that all inflation takes place when a single brane moves down a single throat. In general, there are a number of throats in the bulk in the flux compactification, with different warped factors. As suggested in \cite{Kachru:2002gs}, brane-flux annihilation at the bottom of one throat may lead to a number of $n\ll N_A$ $D$3 branes which can move out of that throat to the bulk. A number of these mobile $D$3-branes move in the bulk and then drop into these throats. So $\phi$ is describing a specific $D$3-brane moving down a specific throat, say, the $A$-throat. After it reaches the bottom, other $D$3-branes are still around and inflation continues. They can generate e-folds before dropping into some other throats, for example in some slow-roll scenarios. This implies that the CMB observables in WMAP correspond to a much smaller number e-folds away from its annihilation with the $\D$3-brane at the bottom of the $A$-throat. As an illustration, suppose there are 50 e-folds coming from other branes, then $\phi$ has to go through only 5 e-folds before its annihilation. 
Following the Lyth bound (\ref{Lyth}), we see that typical $r$ can be 100 times bigger than before, thus allowing a large tensor mode. This also requires a much smaller $\phi$ to begin with, allowing the model to satisfy the bulk volume bound. The predictions at $k_H$ may then depend on more parameters. 

 In fact, inflation can take place when the branes are coming out of a throat, in the case when  
$m^{2} <0$, as in the Chen scenario \cite{Chen:2004gc,Chen:2005ad}. This scenario will easily satisfy the bound (\ref{Liambound}). However, a large tensor mode seems very unlikely since $\gamma$ is decreasing as the brane is coming out of the throat.

$\bullet$ {\it $D$5-, $D$7- and/or $D$9-branes} We have considered only the case of a $D$3-brane moving in the throat. While this may be the simplest and most natural scenario, one may also consider wrapped $D$5- or $D$7-branes. Here, $D$5-branes can wrap a 2-cycle of the throat, while $D$7-branes can wrap any of the 4-cycles that are generically present in the flux compactification. A $D$9-brane wraps both cycles. In these cases, there are additional parameters involved coming from the wrapped cycles. The predictions of a wrapped $D$5-brane depend on a different power of $N_A$ than the $D$3 case. It is easy to see that the change in predictions is related to the different normalization relating the canonical inflaton to the $D$5 position \cite{BLSinprogress}. The $D$5 action is proportional to the $D$5 brane tension, so once the integration over the wrapped two cycle ($\Sigma_2$) is performed, the 4-dimensional action is proportional to $T_5Vol_{\Sigma_2}$. For a particular choice of embedding, one finds $(\Delta\phi/M_p)^2\sim\sqrt{g_s/N_Av}$. Here the orbifolding does not cancel out and a smaller power of $N_A$ appears. In addition, one may add fluxes on the brane which allow the model to match the data with relatively small $N_A$. 

There are some other scenarios which, each by itself, will not generate a large tensor mode, but may be combined with the above scenarios and/or with each other to satisfy the bulk volume bound and enhance the magnitude of the tensor mode.

$\bullet$ {\it Squashed throat} Although the warped geometries that are simplest to construct have a symmetry relating the size of the radial direction to the size of the cross-section, there is no reason in principle not to consider more general possibilities. A ``squashed'' throat would provide a different relationship between the inflaton range and the background charge, changing how the consistency bound is applied.

$\bullet$ {\it Motion at the bottom of the throat}
Recall that the throat is a warped deformed conifold, where the bottom of the throat has the geometry of a finite $S^{3}$ and a shrinking $S^{2}$. For simplicity, the scenario we have discussed so far puts both the $\D$3-brane at the bottom of the throat (at $\phi_{A}$) and the mobile $D$3-brane at the same position in the $S^{3}$. In this case, when they are within a string scale distance (when at $\phi_{E}$, where $\phi_{E} > \phi_{A}$), the open string tachyon mode appears, thus inflation ends before the $D$3-brane has reached the bottom of the throat. If the 2 branes are separated in the finite $S^{3}$, then one can get some number of e-folds after the $D$3-brane has reached the bottom but before it annihilates with the $\D$3-brane. This reduces the number of e-folds $\phi$ has to generate while it is moving down the throat, thus easing the constraint coming from the bulk volume bound.

\section{Conclusions}\label{conclusion}

We have discussed the cosmological predictions for DBI inflation, including the dependence of observed initial power spectra observables on the brane evolution and throat geometry. Three inflationary regimes are of interest, namely the slow-roll, intermediate relativistic and ultra-relativistic regimes, with rich properties including large tensor modes and non-Gaussianity.

After numerically integrating the background and perturbation equations we compare the predicted power spectrum properties to current cosmological CMB, galaxy and supernovae data.  We find that the three regimes can fit the data well although the power spectrum normalization and e-fold constraints severely cut into the full multi-dimensional parameter space. 

The small $m^{2}$ parameter region, which corresponds to a specific model within the usual slow-roll scenario, satisfies the cosmological data quite well predicting a spectral tilt around $n_{s}(k_{pivot}) \simeq0.972$. Its predictions are in line with a typical inflationary scenario. 

Before imposing the bulk volume bound (\ref{Liambound}), we find that the (relatively) large $m^{2}$ region, which corresponds to the DBI ultra-relativistic scenario, is compatible with the present data, in agreement with previous analyses. The potential of having 
a large non-Gaussianity \cite{Alishahiha:2004eh,Shiu:2006kj} with 
distinctive bi-spectrum \cite{Chen:2006nt} and trispectrum \cite{Huang:2006eh} is very encouraging to testing this very stringy scenario. However, imposing the bulk volume bound rules out most if not all of the large $\gamma$ region. That the cosmological data strongly restricts the allowed parameter space and brane inflationary scenarios means that the scenario can be tested; this is very encouraging.

Interestingly, we find that the intermediate $m^{2}$ region, for which a numerical analysis is required, where the tensor mode can be large \cite{Shandera:2006ax} and the $n_t-r$ relation deviates from the usual consistency relation, is also consistent with the data, when one neglects the bulk volume condition (\ref{Liambound}). When this condition is imposed, however, the region is ruled out in the simple scenario. There are a number of ways to vary the model to relax this
bound or the constraint coming from this bound; however, in spite of this, we do not expect the ratio $r$ to be much bigger than a few percent. Two possibilities discussed here are the multi-brane multi-throat scenario and the wrapped $D$5-brane scenario. If $r$ is large enough to be measured, the deviation from the $r-n_{t}$ relation in the slow-roll case will be a fine signature for stringy physics.  

The goal here is not only to find a stringy inflationary scenario that fits the data, but that such a scenario will also predict distinctive stringy signatures that may be detected and measured. Besides cosmic strings, searches for distinct string theory signatures in the slow-roll KKLMMT scenario continues. Recently,  the proposal that the Seiberg duality cascade in the KS throat may lead to observable signatures in the CMB power spectrum \cite{Hailu:2006uj} is encouraging. Features in the warp factor or effects due to the fact that the inflaton is really a six-component field may source non-Gaussianity and slightly alter predictions for the power spectrum \cite{Leblond:2006cc, Chen:2006xj}.

Although the predicted tensor contribution of a few percent is beyond the capabilities of the WMAP satellite, this could well be in the detectable realm for the next generation of CMB instruments focusing on precision measurements of CMB BB polarization modes. Examples are, amongst others, the PLANCK, Clover, Spider, and BICEP projects, which hope to get $\Delta r \sim 0.01-0.05$. As such, the near future holds exciting prospects for better testing the rich properties of brane inflation.

Finally, it is worth mentioning that cosmic strings are generically produced toward the end of brane inflation. We find that the cosmic string tension $\mu$ roughly satisfies $10^{-14}<G\mu<10^{-6}$; its particular value depends on the choice of $m$ and $\lambda$. Roughly speaking, small allowed values of $G\mu$, $10^{-14} \lesssim G\mu \lesssim 10^{-10}$ are accompanied by a red tilt in the power spectrum, and in some cases by large non-Gaussianity. 

\vspace{10mm}

\noindent {\bf \large {Acknowledgments}}

\vspace{5mm}

We thank Daniel Baumann, Cliff Burgess, Xingang Chen, David Chernoff, Gia Dvali, Hassan Firouzjahi, 
Girma Hailu, Nick Jones, Shamit Kachru, Igor Klebanov, Louis Leblond, Liam McAllister, 
Fernando Quevedo, Sash Sarangi, Gary Shiu, Horace Stoica, Eva Silverstein, 
Bret Underwood and Ira Wasserman
for valuable discussions.
The work of RB is supported by the National Science Foundation under grants AST-0607018 and PHY-0555216 and utilizing National 
Supercomputing (NCSA) resources under grant TG-AST060029T.  The work of SS is supported by the DOE under DE-FG02-92ER40699. The work of SHT and JX is supported in part by the National Science Foundation under grant PHY-0355005.

\clearpage

\appendix
\section{The numerical approach to solving the equation of motion} \label{app1}

In DBI inflation, the equations of motion are solved based on the Hamilton-Jacobi approach. The equations we want to solve are \cite{Alishahiha:2004eh},
\begin{eqnarray}
H(\phi)^2 &=& \frac{1}{3M_p^2} [V(\phi)+T(\gamma(\phi)-1)]  \label{eom_H} \\
\gamma(\phi) &=& \sqrt{1+4M_p^4T^{-1}H^{\prime}(\phi)^2} \label{eom_gamma} \\
V(\phi)  &=& \frac{1}{2} m^2 \phi^2 +V_{0}\left(1-\frac{1}{N_{A}}\frac{\phi_A^4}{\phi^4}\frac{(\gamma+1)^2}{4\gamma}\right) \label{V}
\end{eqnarray}
Note that the inflaton equation of motion (\ref{exacteom}) (including the $\ddot \phi$ term) is treated exactly here. In the Hamilton-Jacobi approach, we treat all the variables $H$, $\gamma$, $\dot{\phi}$ as a function of $\phi$, thus we can replace a second order differential equation (\ref{exacteom}) with a group of first order differential equations. The advantage of Hamilton-Jacobi approach is even if the potential is dependent on $\gamma$, which in turn depends on $\dot{\phi}$, one can verify that the approach still leads to the same equation of motion as derived from variation of the action.

The essential idea of the numerical approach is the following,
\begin{enumerate}
\item specify the initial condition at a certain position $\phi_i$; here we choose to start at the edge of the throat $\phi_R \equiv \sqrt{T_3}R$.  We also need to give the initial value of $\dot{\phi_i}$. We always assume that at the start of inflation, the speed of the inflaton is much smaller than the warped speed of light, $\dot{\phi_i}^2 \ll T_3 h(\phi_i)^4$ and $\gamma(\phi_R) \sim 1$.
\item calculate $H'(\phi_R)$ using (\ref{eom_gamma}), 
\begin{equation}
H'(\phi) = \sqrt{\frac{(\gamma^2 - 1)T}{4 M_p^4}}
\end{equation}
From (\ref{phidot}), we know $H' > 0$ when the brane is moving into the throat and $H^\prime < 0$ if the brane moves out.

\item calculate $H(\phi_R)$ using (\ref{eom_H})
\begin{equation}
H(\phi) = \sqrt{\frac{[ V(\phi) + T(\gamma - 1)]}{3 M_p^2} }
\end{equation}

\item using $H^\prime(\phi_R)$ and $H(\phi_R)$, calculate $H(\phi_R - \Delta\phi)$ numerically by the Runga-Kutta method.

\item using the new Hubble parameter at $(\phi_R - \Delta\phi)$, (\ref{eom_H}) can give $\gamma(\phi_R - \Delta\phi)$.
If we use (\ref{V}) and (\ref{eom_H}) together, we can solve $\gamma$ through the equation
\begin{eqnarray}
A\gamma^2 &+& B \gamma + C = 0 \\
A &\equiv& \frac{V_0}{4N_{A}} \frac{\phi_A^4}{\phi^4} - T   \\
B &\equiv& \frac{V_0}{2N_{A}} \frac{\phi_A^4}{\phi^4} - \frac{1}{2}m^2\phi^2 + 3H^2 - V_0 + T \\
C &\equiv& \frac{V_0}{4N_{A}} \frac{\phi_A^4}{\phi^4}
\end{eqnarray}

\item use the new calculated $\gamma(\phi_R - \Delta\phi)$ in step 1 and repeat the whole process to evolve the equation of motion.
\end{enumerate}

When the equation of motion is solved numerically, we can get the numerical dependence of $H$ and $\gamma$ on $\phi$, we can further perform numerical differentiation to get $H'(\phi)$ and $H''(\phi)$ and calculate all the inflationary parameters $\epsilon(\phi)$, $\eta(\phi)$ and $\kappa(\phi)$ as defined in (\ref{params1}).

To compare with observational data, we need to calculate the spectrum of density perturbation at observable scales. The scalar mode spectrum is calculated using (\ref{spectralden}) and the tensor mode using (\ref{tensor}). It's important to note that, because of the speed of sound $1/\gamma$ for the scalar mode, the tensor mode and scalar mode generated at the same $\phi$ during inflation correspond to different scales when they exit the horizon. For scalar mode, $k = aH\gamma$, while for tensor mode $k=aH$. In DBI scenario, generically  $\gamma > 1$, the two scales do not match. In order to calculate the tensor to scalar ratio correctly and compare with experiment, special attention must be paid to calculate the ratio on the same scale $k$, not the same inflaton value $\phi$.  To calculate the corresponding spectral index $n_s$ and $n_t$, we avoid using any analytic formulas but simply apply numerical differentiation to the power spectrum. 

\section{Tying the initial power spectrum to observable scales today} \label{app2}

The first constraint from data is the overall normalization of the scalar density perturbation.  COBE gives us normalization at the comoving scale of the horizon today $k_{H}$ 
\begin{eqnarray*}
k_{H} &\approx& 1.17 \times 10^{-4} Mpc^{-1} \\ 
\delta (k_{H}) &=& (1.7 \pm 0.3) \times 10^{-5}
\end{eqnarray*}
This value has a tilt dependence $\delta^{2}(k) = \delta^{2}(k_{H})\times (k/k_{H})^{n_{s}-1}$ and is also calculated assuming no tensor contribution. 

WMAP reports its results using $A_{s}(k_{pivot})\equiv P_{R}(k_{pivot})$ where $k_{pivot} =0.002 /Mpc$. For our analysis we compare the model to the likelihood space allowed by WMAP 3 year CMB data \cite{Spergel:2006hy,Hinshaw:2006ia,Jarosik:2006ib,Page:2006hz} in combination with the normalization marginalized Sloan Digital Sky Survey LRG matter power spectrum \cite{Tegmark:2006az} and SNLS Supernovae sample \cite{Astier:2005qq}. We compare the DBI models to constraints from observations without imposing the consistency relation,  $n_t(k_{pivot})=-r(k_{pivot})/8$, but imposing a prior on $n_t$, $-0.02<n_t<0.$, for which
\begin{eqnarray} \label{As_exp2}
A_{s}(k_{pivot}) &=& (23.4 ^{+1.5 (2.6)}_{-1.4 (2.5)})\times 10^{-10}\\
n_{s}(k_{pivot}) &=&  0.970^{+0.018 (0.035)}_{-0.019 (0.037)}
\end{eqnarray}
at the 68\%  (and 95\%) c.l. and $r(k_{pivot}) <0.29$ at the 95\% c.l..

At the same scale the COBE and power spectrum normalizations are related by 
\begin{equation}
A_{s}(k_{pivot}) = \frac{25}{4}\delta^{2}(k_{pivot})
\end{equation}
To convert $k_{pivot}$ today to the scale $k_{0}$ at the end of inflation, we need to know the scale factor at the end of inflation $a_{infl}$, $a_{infl} k_{0} = k_{pivot}$. 
We can compare the energy density in radiation today to energy density at reheating in order to obtain $a_{infl}$. 

The radiation density is related to the photon temperature via
\begin{equation}
\rho_{rad}(a) [M L^{-3}] = \frac{\pi^{2}}{30}g_{H}(a)\left(\frac{k_{B}T(a)}{\hbar c}\right)^{3}k_{B}T(a)
\end{equation}
with $g_{H}(a_{infl}) =106.75$ compared to  $g_{H}(a_{today}) =3.36$.

Assuming wholly efficient conversion of inflaton energy into radiation, the radiation density at the end of inflation is given by $\rho_{rad}^{infl} = H(a_{infl})^2$.
Equally today, we have
\begin{eqnarray}
\Omega_{rad}^{0} &=&\Omega_{\gamma}^{0} + N_{\nu} \Omega_{\nu}^{0} \\
&=&(1+N_{\nu}((1+f_{\nu})^{-1}-1)) \Omega_{\gamma}^{0} \nonumber \\
&=& (1+0.68) \times 2.3812\times 10^{-5} h^{-2}\left(\frac{T_{cmb}(K)}{2.7K}\right)^{4} \nonumber \\
&=&4.1812\times 10^{-5} h^{-2}  \nonumber \\
H_{0} &=& 8.7578\times 10^{-61} h  M_{p} \\
\Omega_{rad}^{0} H_{0}^{2} &=&3.2069\times 10^{-120} M_{p}^{2}
\end{eqnarray}
If we compare this to $H(a_{inf})^{2}$ from the numerical calculation then
\begin{eqnarray}
\frac{\rho_{rad}^{infl}}{\Omega_{rad}^{0} H_{0}^{2} } &=& \frac{g_{H}(a_{infl})}{g_{H}(today)}\left(\frac{a_{today}}{a_{infl}}\right)^{4}   \\ 
a_{infl} &=& \left(\frac{106.75}{3.36}\right)^{1/4}\left(\frac{3.2069\times 10^{-120}}{H(a_{infl})^{2}}\right)^{1/4}  \nonumber  \label{ainfl_num}\\
&=& \frac{3.17708 \times 10^{-30}}{\sqrt{H(a_{infl})}}
\end{eqnarray}
Once we have $a_{infl}$, we are able to calculate the scale $k_{0}$ that will become the pivot scale $k_{pivot}$ today. All the observables at the pivot scale are generated at $\phi_{pivot}$ during inflation. To calculate $\phi_{pivot}
$, we use the horizon crossing condition
\begin{equation}
c_s k_{0} = a(\phi_{pivot}) H(\phi_{pivot}) \label{hcross}
\end{equation}
Combining (\ref{ainfl_num}) and (\ref{hcross}), we get the minimum number of e-folds to solve the horizon problem as
\begin{equation}\label{min_efold}
N_e = 68.6 + \ln \left(\frac{H_I \gamma}{M_p} \right) - \ln \left(\frac{T_{RH}}{M_p} \right),
\end{equation}
where $H_I$ is the inflation scale and $T_{RH}$ is the reheating temperature, and we assume 100\% efficiency in converting the inflaton energy into radiation.

We numerically interpolate at $\phi_{pivot}$ and calculate the magnitude of scalar density perturbation there. For models which can obtain the normalization requirements for the WMAP+ SDSS LRG+ SNLS SN1a constraint (\ref{As_exp2}) at the 95\% confidence level, we consider the behavior of the scalar spectral index, $n_s$, tensor spectral index, $n_t$, and tensor to scalar ratio, $r$, at $\phi_{pivot}$. 

\clearpage

\bibliographystyle{JHEP}
\bibliography{paper}

\providecommand{\href}[2]{#2}\begingroup\raggedright\begin{thebibliography}{10}

\bibitem{Guth:1980zm}
A.~H. Guth, {\it The inflationary universe: A possible solution to the horizon
  and flatness problems},  {\em Phys. Rev.} {\bf D23} (1981) 347--356.

\bibitem{Linde:1981mu}
A.~D. Linde, {\it A new inflationary universe scenario: A possible solution of
  the horizon, flatness, homogeneity, isotropy and primordial monopole
  problems},  {\em Phys. Lett.} {\bf B108} (1982) 389--393.

\bibitem{Albrecht:1982wi}
A.~Albrecht and P.~J. Steinhardt, {\it Cosmology for grand unified theories
  with radiatively induced symmetry breaking},  {\em Phys. Rev. Lett.} {\bf 48}
  (1982) 1220--1223.

\bibitem{Smoot:1992td}
G.~F. Smoot {\em et~al.}, {\it Structure in the cobe dmr first year maps},
  {\em Astrophys. J.} {\bf 396} (1992) L1--L5.

\bibitem{Spergel:2006hy}
D.~N. Spergel {\em et~al.}, {\it Wilkinson microwave anisotropy probe (wmap)
  three year results: Implications for cosmology},
  \href{http://xxx.lanl.gov/abs/astro-ph/0603449}{{\tt astro-ph/0603449}}.

\bibitem{Dvali:1998pa}
G.~R. Dvali and S.~H.~H. Tye, {\it Brane inflation},  {\em Phys. Lett.} {\bf
  B450} (1999) 72--82, [\href{http://xxx.lanl.gov/abs/hep-ph/9812483}{{\tt
  hep-ph/9812483}}].

\bibitem{Kachru:2003sx}
S.~Kachru, R.~Kallosh, A.~Linde, J.~Maldacena, L.~McAllister, and S.~P.
  Trivedi, {\it Towards inflation in string theory},  {\em JCAP} {\bf 0310}
  (2003) 013, [\href{http://xxx.lanl.gov/abs/hep-th/0308055}{{\tt
  hep-th/0308055}}].

\bibitem{Tye:2006uv}
S.~H.~H. Tye, {\it Brane inflation: String theory viewed from the cosmos},
  \href{http://xxx.lanl.gov/abs/hep-th/0610221}{{\tt hep-th/0610221}}.

\bibitem{Hinshaw:2006ia}
G.~Hinshaw {\em et~al.}, {\it Three-year wilkinson microwave anisotropy probe
  (wmap) observations: Temperature analysis},
  \href{http://xxx.lanl.gov/abs/astro-ph/0603451}{{\tt astro-ph/0603451}}.

\bibitem{Jarosik:2006ib}
N.~Jarosik {\em et~al.}, {\it Three-year wilkinson microwave anisotropy probe
  (wmap) observations: Beam profiles, data processing, radiometer
  characterization and systematic error limits},
  \href{http://xxx.lanl.gov/abs/astro-ph/0603452}{{\tt astro-ph/0603452}}.

\bibitem{Page:2006hz}
L.~Page {\em et~al.}, {\it Three year wilkinson microwave anisotropy probe
  (wmap) observations: Polarization analysis},
  \href{http://xxx.lanl.gov/abs/astro-ph/0603450}{{\tt astro-ph/0603450}}.

\bibitem{Tegmark:2006az}
M.~Tegmark {\em et~al.}, {\it Cosmological constraints from the sdss luminous
  red galaxies},  {\em Phys. Rev.} {\bf D74} (2006) 123507,
  [\href{http://xxx.lanl.gov/abs/astro-ph/0608632}{{\tt astro-ph/0608632}}].

\bibitem{Astier:2005qq}
P.~Astier {\em et~al.}, {\it The supernova legacy survey: Measurement of
  $\omega_m$, $\omega_\lambda$ and w from the first year data set},  {\em
  Astron. Astrophys.} {\bf 447} (2006) 31--48,
  [\href{http://xxx.lanl.gov/abs/astro-ph/0510447}{{\tt astro-ph/0510447}}].

\bibitem{Silverstein:2003hf}
E.~Silverstein and D.~Tong, {\it Scalar speed limits and cosmology:
  Acceleration from d- cceleration},
  \href{http://xxx.lanl.gov/abs/hep-th/0310221}{{\tt hep-th/0310221}}.

\bibitem{Firouzjahi:2005dh}
H.~Firouzjahi and S.~H.~H. Tye, {\it Brane inflation and cosmic string tension
  in superstring theory},  {\em JCAP} {\bf 0503} (2005) 009,
  [\href{http://xxx.lanl.gov/abs/hep-th/0501099}{{\tt hep-th/0501099}}].

\bibitem{Seljak:2006hi}
U.~Seljak and A.~Slosar, {\it B polarization of cosmic microwave background as
  a tracer of strings},  {\em Phys. Rev.} {\bf D74} (2006) 063523,
  [\href{http://xxx.lanl.gov/abs/astro-ph/0604143}{{\tt astro-ph/0604143}}].

\bibitem{Shandera:2006ax}
S.~E. Shandera and S.~H.~H. Tye, {\it Observing brane inflation},
  \href{http://xxx.lanl.gov/abs/hep-th/0601099}{{\tt hep-th/0601099}}.

\bibitem{Alishahiha:2004eh}
M.~Alishahiha, E.~Silverstein, and D.~Tong, {\it Dbi in the sky},  {\em Phys.
  Rev.} {\bf D70} (2004) 123505,
  [\href{http://xxx.lanl.gov/abs/hep-th/0404084}{{\tt hep-th/0404084}}].

\bibitem{Chen:2006hs}
X.~Chen, S.~Sarangi, S.~H. Henry~Tye, and J.~Xu, {\it Is brane inflation
  eternal?},  {\em JCAP} {\bf 0611} (2006) 015,
  [\href{http://xxx.lanl.gov/abs/hep-th/0608082}{{\tt hep-th/0608082}}].

\bibitem{Baumann:2006th}
D.~Baumann, A.~Dymarsky, I.~R. Klebanov, J.~Maldacena, L.~McAllister, and
  A.~Murugan, {\it On d3-brane potentials in compactifications with fluxes and
  wrapped d-branes},  \href{http://xxx.lanl.gov/abs/hep-th/0607050}{{\tt
  hep-th/0607050}}.

\bibitem{Giddings:2001yu}
S.~B. Giddings, S.~Kachru, and J.~Polchinski, {\it Hierarchies from fluxes in
  string compactifications},  {\em Phys. Rev.} {\bf D66} (2002) 106006,
  [\href{http://xxx.lanl.gov/abs/hep-th/0105097}{{\tt hep-th/0105097}}].

\bibitem{Kachru:2003aw}
S.~Kachru, R.~Kallosh, A.~Linde, and S.~P. Trivedi, {\it De sitter vacua in
  string theory},  {\em Phys. Rev.} {\bf D68} (2003) 046005,
  [\href{http://xxx.lanl.gov/abs/hep-th/0301240}{{\tt hep-th/0301240}}].

\bibitem{Burgess:2001fx}
C.~P. Burgess, M.~Majumdar, D.~Nolte, F.~Quevedo, G.~Rajesh, and R.~Zhang, {\it
  The inflationary brane-antibrane universe},  {\em JHEP} {\bf 07} (2001) 047,
  [\href{http://xxx.lanl.gov/abs/hep-th/0105204}{{\tt hep-th/0105204}}].

\bibitem{Dvali:2001fw}
G.~R. Dvali, Q.~Shafi, and S.~Solganik, {\it D-brane inflation},
  \href{http://xxx.lanl.gov/abs/hep-th/0105203}{{\tt hep-th/0105203}}.

\bibitem{Shandera:2003gx}
S.~Shandera, B.~Shlaer, H.~Stoica, and S.~H.~H. Tye, {\it Inter-brane
  interactions in compact spaces and brane inflation},  {\em JCAP} {\bf 0402}
  (2004) 013, [\href{http://xxx.lanl.gov/abs/hep-th/0311207}{{\tt
  hep-th/0311207}}].

\bibitem{Fradkin:1985qd}
E.~S. Fradkin and A.~A. Tseytlin, {\it Nonlinear electrodynamics from quantized
  strings},  {\em Phys. Lett.} {\bf B163} (1985) 123.

\bibitem{Abouelsaood:1986gd}
A.~Abouelsaood, J.~Callan, Curtis~G., C.~R. Nappi, and S.~A. Yost, {\it Open
  strings in background gauge fields},  {\em Nucl. Phys.} {\bf B280} (1987)
  599.

\bibitem{Kehagias:1999vr}
A.~Kehagias and E.~Kiritsis, {\it Mirage cosmology},  {\em JHEP} {\bf 11}
  (1999) 022, [\href{http://xxx.lanl.gov/abs/hep-th/9910174}{{\tt
  hep-th/9910174}}].

\bibitem{Burgess:2003qv}
C.~P. Burgess, P.~Martineau, F.~Quevedo, and R.~Rabadan, {\it Branonium},  {\em
  JHEP} {\bf 06} (2003) 037,
  [\href{http://xxx.lanl.gov/abs/hep-th/0303170}{{\tt hep-th/0303170}}].

\bibitem{Polchinski:1998rr}
J.~Polchinski, {\em String Theory. Vol. 2: Superstring Theory and Beyond}.
\newblock Cambridge Univ. Pr., 1998.

\bibitem{Gubser:1998vd}
S.~S. Gubser, {\it Einstein manifolds and conformal field theories},  {\em
  Phys. Rev.} {\bf D59} (1999) 025006,
  [\href{http://xxx.lanl.gov/abs/hep-th/9807164}{{\tt hep-th/9807164}}].

\bibitem{McAllister:2005mq}
L.~McAllister, {\it An inflaton mass problem in string inflation from threshold
  corrections to volume stabilization},  {\em JCAP} {\bf 0602} (2006) 010,
  [\href{http://xxx.lanl.gov/abs/hep-th/0502001}{{\tt hep-th/0502001}}].

\bibitem{Shandera:2004zy}
S.~E. Shandera, {\it Slow roll in brane inflation},  {\em JCAP} {\bf 0504}
  (2005) 011, [\href{http://xxx.lanl.gov/abs/hep-th/0412077}{{\tt
  hep-th/0412077}}].

\bibitem{Berg:2004ek}
M.~Berg, M.~Haack, and B.~Kors, {\it Loop corrections to volume moduli and
  inflation in string theory},  {\em Phys. Rev.} {\bf D71} (2005) 026005,
  [\href{http://xxx.lanl.gov/abs/hep-th/0404087}{{\tt hep-th/0404087}}].

\bibitem{Berg:2004sj}
M.~Berg, M.~Haack, and B.~Kors, {\it On the moduli dependence of
  nonperturbative superpotentials in brane inflation},
  \href{http://xxx.lanl.gov/abs/hep-th/0409282}{{\tt hep-th/0409282}}.

\bibitem{Balasubramanian:2004uy}
V.~Balasubramanian and P.~Berglund, {\it Stringy corrections to kaehler
  potentials, susy breaking, and the cosmological constant problem},  {\em
  JHEP} {\bf 11} (2004) 085,
  [\href{http://xxx.lanl.gov/abs/hep-th/0408054}{{\tt hep-th/0408054}}].

\bibitem{Burgess:2006cb}
C.~P. Burgess, J.~M. Cline, K.~Dasgupta, and H.~Firouzjahi, {\it Uplifting and
  inflation with d3 branes},
  \href{http://xxx.lanl.gov/abs/hep-th/0610320}{{\tt hep-th/0610320}}.

\bibitem{Kabat:1999yq}
D.~Kabat and G.~Lifschytz, {\it Gauge theory origins of supergravity causal
  structure},  {\em JHEP} {\bf 05} (1999) 005,
  [\href{http://xxx.lanl.gov/abs/hep-th/9902073}{{\tt hep-th/9902073}}].

\bibitem{Chen:2006nt}
X.~Chen, M.-x. Huang, S.~Kachru, and G.~Shiu, {\it Observational signatures and
  non-gaussianities of general single field inflation},
  \href{http://xxx.lanl.gov/abs/hep-th/0605045}{{\tt hep-th/0605045}}.

\bibitem{Baumann:2006cd}
D.~Baumann and L.~McAllister, {\it A microscopic limit on gravitational waves
  from d-brane inflation},  \href{http://xxx.lanl.gov/abs/hep-th/0610285}{{\tt
  hep-th/0610285}}.

\bibitem{Garriga:1999vw}
J.~Garriga and V.~F. Mukhanov, {\it Perturbations in k-inflation},  {\em Phys.
  Lett.} {\bf B458} (1999) 219--225,
  [\href{http://xxx.lanl.gov/abs/hep-th/9904176}{{\tt hep-th/9904176}}].

\bibitem{Stewart:1993bc}
E.~D. Stewart and D.~H. Lyth, {\it A more accurate analytic calculation of the
  spectrum of cosmological perturbations produced during inflation},  {\em
  Phys. Lett.} {\bf B302} (1993) 171--175,
  [\href{http://xxx.lanl.gov/abs/gr-qc/9302019}{{\tt gr-qc/9302019}}].

\bibitem{Creminelli:2005hu}
P.~Creminelli, A.~Nicolis, L.~Senatore, M.~Tegmark, and M.~Zaldarriaga, {\it
  Limits on non-gaussianities from wmap data},  {\em JCAP} {\bf 0605} (2006)
  004, [\href{http://xxx.lanl.gov/abs/astro-ph/0509029}{{\tt
  astro-ph/0509029}}].

\bibitem{Huang:2006eh}
M.-x. Huang and G.~Shiu, {\it The inflationary trispectrum for models with
  large non- gaussianities},
  \href{http://xxx.lanl.gov/abs/hep-th/0610235}{{\tt hep-th/0610235}}.

\bibitem{Shiu:2006kj}
G.~Shiu and B.~Underwood, {\it Observing the geometry of warped
  compactification via cosmic inflation},
  \href{http://xxx.lanl.gov/abs/hep-th/0610151}{{\tt hep-th/0610151}}.

\bibitem{Kecskemeti:2006cg}
S.~Kecskemeti, J.~Maiden, G.~Shiu, and B.~Underwood, {\it Dbi inflation in the
  tip region of a warped throat},  {\em JHEP} {\bf 09} (2006) 076,
  [\href{http://xxx.lanl.gov/abs/hep-th/0605189}{{\tt hep-th/0605189}}].

\bibitem{Klebanov:2000hb}
I.~R. Klebanov and M.~J. Strassler, {\it Supergravity and a confining gauge
  theory: Duality cascades and chisb-resolution of naked singularities},  {\em
  JHEP} {\bf 08} (2000) 052,
  [\href{http://xxx.lanl.gov/abs/hep-th/0007191}{{\tt hep-th/0007191}}].

\bibitem{Chen:2004gc}
X.-g. Chen, {\it Multi-throat brane inflation},
  \href{http://xxx.lanl.gov/abs/hep-th/0408084}{{\tt hep-th/0408084}}.

\bibitem{Chen:2005ad}
X.-g. Chen, {\it Inflation from warped space},  {\em JHEP} {\bf 08} (2005) 045,
  [\href{http://xxx.lanl.gov/abs/hep-th/0501184}{{\tt hep-th/0501184}}].

\bibitem{Kachru:2002gs}
S.~Kachru, J.~Pearson, and H.~L. Verlinde, {\it Brane/flux annihilation and the
  string dual of a non- supersymmetric field theory},  {\em JHEP} {\bf 06}
  (2002) 021, [\href{http://xxx.lanl.gov/abs/hep-th/0112197}{{\tt
  hep-th/0112197}}].

\bibitem{Chen:2005fe}
X.~Chen, {\it Running non-gaussianities in dbi inflation},  {\em Phys. Rev.}
  {\bf D72} (2005) 123518,
  [\href{http://xxx.lanl.gov/abs/astro-ph/0507053}{{\tt astro-ph/0507053}}].

\bibitem{Chen:2006ni}
X.~Chen and S.~H.~H. Tye, {\it Heating in brane inflation and hidden dark
  matter},  {\em JCAP} {\bf 0606} (2006) 011,
  [\href{http://xxx.lanl.gov/abs/hep-th/0602136}{{\tt hep-th/0602136}}].

\bibitem{Hailu:2006uj}
G.~Hailu and S.~H.~H. Tye, {\it Structures in the gauge / gravity duality
  cascade},  \href{http://xxx.lanl.gov/abs/hep-th/0611353}{{\tt
  hep-th/0611353}}.

\bibitem{Lewis:2002ah}
A.~Lewis and S.~Bridle, {\it Cosmological parameters from cmb and other data: a
  monte- carlo approach},  {\em Phys. Rev.} {\bf D66} (2002) 103511,
  [\href{http://xxx.lanl.gov/abs/astro-ph/0205436}{{\tt astro-ph/0205436}}].

\bibitem{BLSinprogress}
M.~Becker, L.~Leblond, and S.~Shandera, {\it In progress}, .

\bibitem{Leblond:2006cc}
L.~Leblond and S.~Shandera, {\it Cosmology of the tachyon in brane inflation},
  \href{http://xxx.lanl.gov/abs/hep-th/0610321}{{\tt hep-th/0610321}}.

\bibitem{Chen:2006xj}
X.~Chen, R.~Easther, and E.~A. Lim, {\it Large non-gaussianities in single
  field inflation},  \href{http://xxx.lanl.gov/abs/astro-ph/0611645}{{\tt
  astro-ph/0611645}}.

\end{thebibliography}\endgroup

\end{document}